\documentclass[aps, pra, preprint, groupedaddress, amsfonts,
               amsmath, amssymb, showpacs, nofootinbib]{revtex4-1}
\usepackage{microtype}
\usepackage{graphicx}
\usepackage{epstopdf}
\usepackage[utf8]{inputenc}
\usepackage[T1]{fontenc}
\usepackage[usenames,dvipsnames]{xcolor}
\usepackage{hyperref}

\usepackage{color}% \textcolor{#1}{#2}

\begin{document}
\title{Non-perturbative scaling behavior for net ionization of biologically relevant molecules by multiply-charged heavy-ion impact}

\author{Hans J\"urgen L\"udde}
\email[]{luedde@itp.uni-frankfurt.de}
\affiliation{Frankfurt Institute for Advanced Studies (FIAS), D-60438 Frankfurt, Germany} 

\author{Thilo Kalkbrenner}
\email[]{kalkbrenner@csc.uni-frankfurt.de}
\affiliation{Center for Scientific Computing, Goethe-Universit\"at, D-60438 Frankfurt, Germany}

\author{Marko Horbatsch}
\email[]{marko@yorku.ca}
\affiliation{Department of Physics and Astronomy, York University, Toronto, Ontario M3J 1P3, Canada}

\author{Tom Kirchner}  
\email[]{tomk@yorku.ca}
\affiliation{Department of Physics and Astronomy, York University, Toronto, Ontario M3J 1P3, Canada}
\date{\today}
\begin{abstract}
A recently developed model to describe proton collisions from molecules involving basic atoms such as hydrogen, carbon, nitrogen, oxygen and phosphorus (H, C, N, O, P)
is extended to treat collisions with multiply charged ions. The ion-atom collisions are computed using the two-center basis generator method (TC-BGM), 
which has a proven track record of 
yielding accurate total cross sections for electron capture and ionization. The atomic net ionization cross sections are then used to assemble two models for ion-molecule collisions: an independent atom model (IAM) that follows the Bragg additivity rule (labeled IAM-AR), and also the so-called pixel-counting method (IAM-PCM). The latter yields reduced cross sections relative to IAM-AR near the maximum, since it takes into account the overlapping nature of effective cross sectional areas. The IAM-PCM for higher-charge projectiles leads to strong reductions of net ionization cross sections relative to the IAM-AR method, and is computed directly for projectile charges $Q=1, 2, 3$. The scaling behavior of the IAM-PCM is investigated over a wide range of energies $E$, and at high $E$ it converges towards the IAM-AR. An empirical scaling rule based on the IAM-PCM results is established which allows to reproduce these results based on proton impact calculations. Detailed comparisons are provided for the uracil target 
($\rm C_4 H_4 N_2 O_2$), for which other theoretical as well as experimental results are available. Data are also shown for targets such as water ($\rm H_2 O$), methane ($\rm CH_4$),
adenine ($\rm C_5 H_5 N_5 $), L-valine ($\rm C_5 H_{11} N O_2$), and the nucleotide dAMP ($\rm C_{10} H_{14} N_5 O_6 P$).
Based on the scaling model derived from the IAM-PCM cross sections it is shown how the experimental data for uracil and water bombarded by multiply charged ions can be reduced to
effective $Q=1$ cross sections respectively, and these are compared to proton impact data.
\end{abstract}
%
%\pacs{34.10.+x, 34.50.Gb, 34.70.+e, 36.40.-c}
%

\maketitle
\section{Introduction}
\label{intro}
The understanding of ionization in charged-particle impact on biologically relevant molecules is an important prerequisite for ion beam cancer therapy.
In order to perform simulations of radiation damage caused by the projectile ions, the secondary electrons and molecular charged fragments 
produced during collisions~\cite{Alcocer2019}, one first has to study
the fundamental processes of ion collisions with molecules in the gas/vapor phase, and this has been the motivation for experimental and theoretical studies.
Important target molecules in this context are water (for which theoretical studies can be performed as extensions of ion-atom collisions), and biomolecules that form
the DNA and RNA. The RNA base uracil ($\rm C_4 H_4 N_2 O_2$) was chosen as a candidate for extensive experimentation and was reported on together with theoretical 
analyses~\cite{PhysRevA.85.032711,Agnihotri_2013}. 
Theoretical treatment of ionization of these biomolecules escapes
at this point the capabilities of sophisticated quantum-mechanical modelling, but first attempts have been made~\cite{PhysRevLett.107.023202,Covington17,Salo18}. The net ionization cross sections for biomolecules
are very large due to their size and number of available valence electrons.

In Ref.~\cite{Itoh13} an experimental (and theoretical) investigation of p-uracil collisions summarizes how the ionization cross section grows with the number of valence 
electrons. This idea also forms the basis of a theoretical approach that combines ion-atom scattering calculations within the continuum distorted wave with eikonal initial state
approach (CDW-EIS), which also uses information from the molecular orbital energy-level structure to imprint some molecular character on the model. Another approach
that was used with some success is an independent-molecule model, where parametrizations of experimental total ionization cross section data for proton 
collisions with small constituent molecules are used to assemble results for the uracil target~\cite{Paredes15}.

Independent-atom models
(IAM) were studied extensively for proton impact in our group~\cite{hjl16,hjl18,hjl19,hjl19b}. 
On the basis of proton-atom collision calculations performed using the two-center basis generator method (TC-BGM)~\cite{tcbgm} 
one can form simply an estimate for the ion-molecule cross section by the Bragg additivity rule, and this was called the IAM-AR model. This somewhat naive addition rule
should provide the correct high-impact-energy limit when the ionization cross section becomes small. A much more sophisticated IAM was introduced and tested in these 
works, which models the ion-molecule collision process for a given orientation of the molecule by considering the projectile ion as observing an effective cross sectional area
that is formed as the overlap of all atomic cross sections. This geometric overlap is easily calculated by pixelization, and therefore the method was called the pixel counting
method (PCM). This method gives substantially reduced cross sections for net ionization and net capture as compared to the IAM-AR method, but goes over into it by
construction when the cross sections become small and the overlap disappears, i.e., the measured area corresponds to the sum of all atomic cross sections.
Quite relevant for the present work is the fact that the IAM-PCM results for net ionization in p-uracil collisions at the energies $E=0.5, 1.0, 2.0 \ \rm MeV$ match the experimental data
quite well, as does the CDW-EIS theory for proton impact~\cite{Itoh13}. In Ref.\cite{hjl19} a detailed comparison between IAM-AR, IAM-PCM and 
molecular-orbital energy based models is provided for proton-pyrimidine $\rm (C_4 H_4 N_2)$ collisions in Fig. 2a, and it is evident
that the shape of the IAM-PCM cross sections begins to deviate at energies below $E=1 \ \rm MeV/amu$ from AR and the other models. A comparison for the case of adenine molecules
is provided in Fig.~3 of Ref.~\cite{hjl19b}.

An important measure that helps to understand how theory and experiment match up is the scaling behavior of the ionization cross sections with projectile charge $Q$
and collision energy $E$ or impact velocity $v$. An experimental (and theoretical) investigation of charged-ion collisions with uracil~\cite{Agnihotri_2013} found that even in 
the MeV/amu collision energy regime perturbative, and even distorted-wave theory are not in good agreement. It is argued that the experimental cross 
section scales approximately as $\eta^{1.5}$, where $\eta=Q/v$ is the Sommerfeld parameter. 
The CDW-EIS theory overestimates the experimental data by more than a factor of two. Thus, it is important to investigate
the situation using a theory that is not rooted in the Born series or its distorted-wave siblings. 
A recent IAM based on a stoichiometric argument and CDW-EIS theory for ion-atom collisions~\cite{mendez2019ionization}
makes an attempt at scaling by defining a reduced CDW-EIS cross section, which however fails to reconcile the proton impact measurements~\cite{Itoh13} with the $Q=4-8$
data of Ref.~\cite{Agnihotri_2013} with a difference of about a factor of three (cf. Fig.~5 in Ref.~\cite{mendez2019ionization}). This problem shows that the scaling problem
has remained unresolved, and it therefore represents a major thrust for the present work. We also note that the follow-up work from the CDW-EIS scaling approach~\cite{alej2020universal} is not attempting to reconcile their findings with all measured uracil data.

A significant body of ionization data for atoms and molecules, including some biomolecules has been treated theoretically by a semiclassical method that represents
an $\hbar=0$ limit of quantum mechanics, namely the Classical Trajectory Monte Carlo (CTMC) method which simulates quantum mechanics by classical statistical mechanics. 
It allows to obtain differential electron emission cross sections and is non-perturbative. Molecules can be described by multi-center effective model potentials adopted from
quantum structure calculations. For the water molecule target bombarded by highly charged ions a detailed comparison against CDW-EIS results and experiments was
carried out recently in a model that includes time-dependent screening parameters that change with electron removal from the molecule~\cite{PhysRevA.99.062701}.

Experimental data for charged-particle impact with $Q=1,2,3$ on water molecules have been obtained for a wide range of collision energies~\cite{Toburen_80,PhysRevA.32.2128,Bolorizadeh86,Luna07,Luna16}, and for the remainder of the Introduction we focus on this target molecule. 
For higher charges $Q$ measurements are available at higher collision 
energies~\cite{Ohsawa_2013,Bhattacharjee_2016,Bhattacharjee_2017,Bhattacharjee_2018} 
in the form of differential electron emission cross sections together with CDW-EIS theory, and integration of the differential data determines total ionization cross sections.

Calculations based on the TC-BGM were carried out for proton-$\rm H_2O$ collisions by representing the self-consistent field molecular orbitals in terms of an
expanded atomic oxygen basis generated by the optimized potential method of density functional theory~\cite{hjl09}. Not all molecular orientations could be 
treated, but arguments were given how the total cross sections could be estimated on the basis of a subset of orientations. The method was applied successfully 
to the cases of proton projectiles~\cite{PhysRevA.85.052704} and a fragmentation model was developed and compared to experimental data~\cite{PhysRevA.85.052713}.
The work was extended to $\rm He^{+}$ collisions including the treatment of projectile electron loss~\cite{PhysRevA.86.022719}. 
The experimental work on $\rm Li^{3+}$ collisions was supported using this methodology~\cite{Luna16}.
However, the extension to higher multiply charged projectiles turned out to be difficult (despite some additional encouraging results for $\rm He^{2+}$ projectiles~\cite{Pausz_2014}), 
and the program was not pursued further in favor of the IAM-PCM approach. The main motivation for this was that TC-BGM ion-atom calculations are pushed
to convergence including a proper representation of the discrete excited states of the constituent atoms. 
This turns out to be difficult for the ion-molecule TC-BGM calculations, in part due to the representation in terms of atomic oxygen basis states. A proper representation
of the molecular discrete excitation spectrum requires a correlated quantum chemistry approach~\cite{doi:10.1063/1.2837827}. Thus, it is not clear whether a
limited representation within the framework of the TC-BGM ion-molecule calculations is capable of separating accurately the continuum contributions from discrete
excitations.

Other groups tested methods based on the semiclassical approach
where the electronic wave function developed in a molecular orbital basis is propagated in time~\cite{PhysRevA.87.032709} and obtained different (higher) net ionization 
cross sections at low energies.  A full three-center  single-electron CTMC model
potential was designed~\cite{PhysRevA.83.052704} that gained popularity within the community with extensive sampling of
molecular orientations. The potential was also used in
numerical grid calculations~\cite{ERREA2015} at higher collision energies.

An attempt at scaling was recently made in the context of CTMC calculations~\cite{Otranto2019} by graphing the net ionization cross section divided by the projectile charge $Q$
against $E'=E/Q$. The CTMC data run approximately parallel to the CDW-EIS data and experimental data  at high energies, and begin to deviate from experimental data
as they approach $E'=100 \ \rm keV/amu$. from above. The question whether the cross section data can be scaled within the maximum region therefore also remains open
up to now.

A recently introduced CTMC mean field model with dynamic screening on projectile and target has been used to make extensive comparisons with the data and explored the question
of saturation behavior in the net cross sections by looking at them as a function of the Sommerfeld parameter $\eta=Q/v$~\cite{jorge2020multicharged}. No obvious scaling was observed in these calculations when looking at the entire energy range in the data.

The paper is organized as follows. In Sect.~\ref{sec:model} we introduce the theoretical basis for the current work. Sect.~\ref{sec:model1} presents new results for
TC-BGM ion-atom calculations for $\rm He^{2+}$ and $\rm Li^{3+}$ projectiles; for atomic hydrogen targets the results are compared with theory and experiment.
In Sect.~\ref{sec:model2} we demonstrate the scaling behavior of ion-molecule collisions with respect to projectile charge $Q$ and collision energy $E$ in the IAM-AR and IAM-PCM approaches for the targets uracil and water. A parametrization of the scaling behavior is then used to generate cross section for arbitrary projectile charge on the basis
of %IAM-AR and% 
IAM-PCM proton impact data. Sect.~\ref{sec:expt} serves to provide a detailed comparison of the scaling behavior: in Sect.~\ref{sec:expt1} for uracil
and in Sect.~\ref{sec:expt2} for water. The paper ends with a few concluding remarks in Sect.~\ref{sec:conclusions}.
Atomic units, characterized by $\hbar=m_e=e=4\pi\epsilon_0=1$, are used unless otherwise stated.

\section{Model}
\label{sec:model}
In this section we present arguments that lead to the
theoretical data-driven empirical scaling models for the net ionization cross sections according to the IAM-AR and IAM-PCM models using the uracil target molecule $\rm C_4 H_4 N_2 O_2$ as an example. It is a good example from several points of view: on the one hand it contains four different atoms, and ionization shouldn't be dominated by any one of them, and on the other hand it is a target for which systematic experimental studies in the gas phase were carried out. The model will then later be confirmed by looking at the water molecule (vapor phase), for which again, a number of experiments with differently charged projectile ions are available.

\subsection{Ion-atom collisions}
\label{sec:model1}

Results from new TC-BGM solutions of the time-dependent Schr\"odinger equation 
(reduced to density functional theory at the level of an independent electron model) are presented here 
for collisions of multi-charged ions (bare charges $Q=1,2,3$) from atoms that form constituents of the
biologically relevant molecules, i.e., (H, C, N, O, P). 
The case of proton projectiles was treated explicitly in Ref.~\cite{hjl19}, so those results are not repeated here. 
The TC-BGM is implemented in projectile potential $W_{\rm P}$ hierarchy which converges well and adds 123 pseudostates to span the complementary 
space (for all projectile and
target combinations) to the number of shells treated explicitly for target and projectile. For the target the $n_{\rm T} = 1, ...,5$ shells are included in all cases.
For the projectile we include $n_{\rm P} = 1, ..., 6$ in the case of $\rm He^{2+}$ projectiles and $n_{\rm P} = 1, ..., 7$ for $\rm Li^{3+}$.
For ionization we observe a simple trend
for high collision energies, namely $\sigma_{\rm C}=\sigma_{\rm N}=\sigma_{\rm O} = 4 \sigma_{\rm H} = 2 \sigma_{\rm P}/3$, as was found previously for proton impact~\cite{hjl19}.

Pilot calculations were also carried out for $Q=6$ projectiles in order to test the scaling predictions. These TC-BGM ion-atom calculations for larger $Q$
require order-of-magnitude increases in computer time, since one has to take into account capture into shells with high principal quantum numbers. This increase in 
demand on computer resources is one of the primary motivations for the present work that involves the scaling of IAM-PCM data using $Q=2,3$ projectiles, relating
them to the $Q=1$ calculations, and then extrapolating to higher $Q$. Currently, there is access to experimental data for projectiles charges 
as high as $Q=13$ for uracil and water targets, and these data are used in the present work for validation.
Whether the extrapolations can be trusted to higher $Q$ values remains to be seen once more experimental data become available.

\begin{figure}
\begin{center}$
%\resizebox{0.6\textwidth}{!}{%
\begin{array}{cc}
\resizebox{0.5\textwidth}{!}{\includegraphics{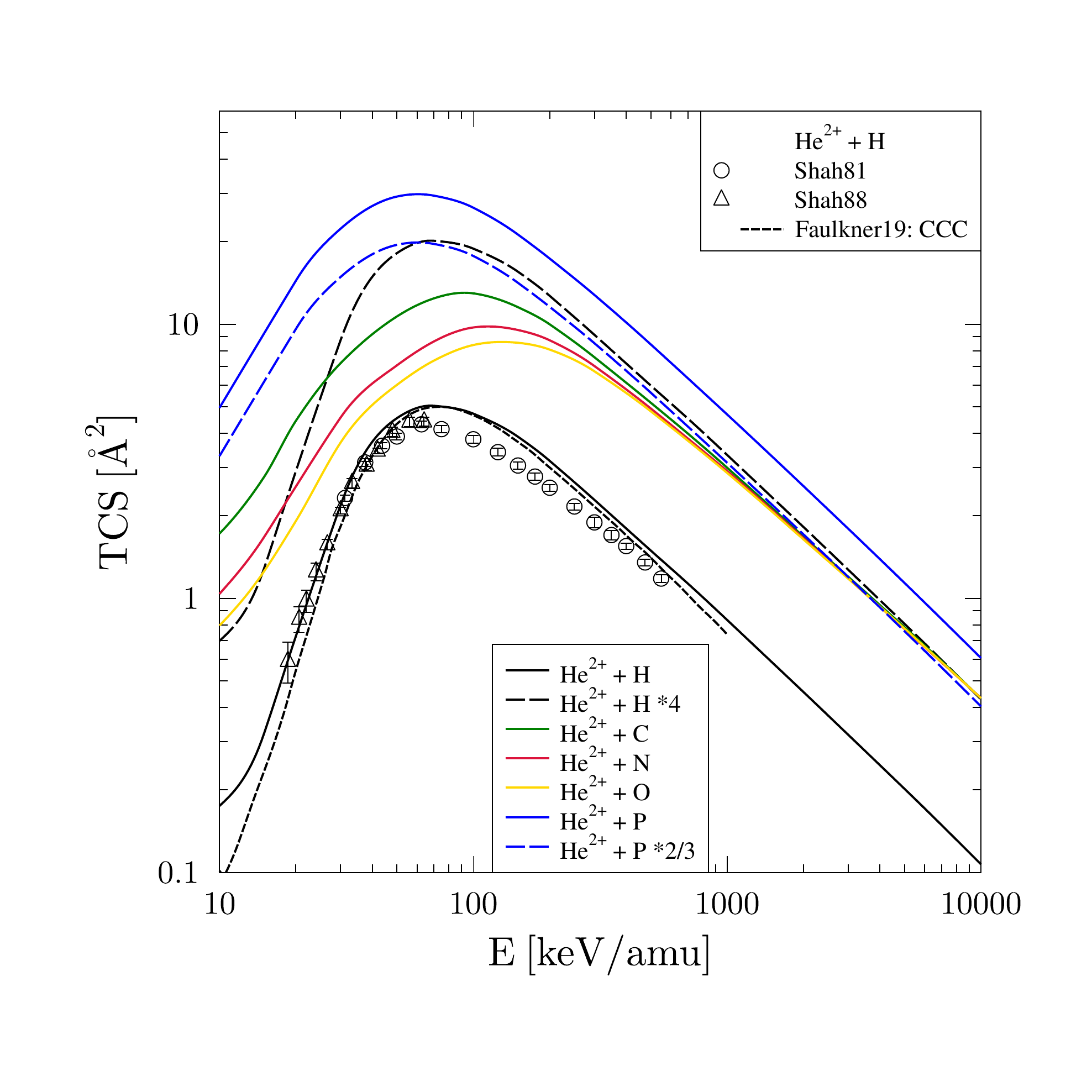}}&
\resizebox{0.5\textwidth}{!}{\includegraphics{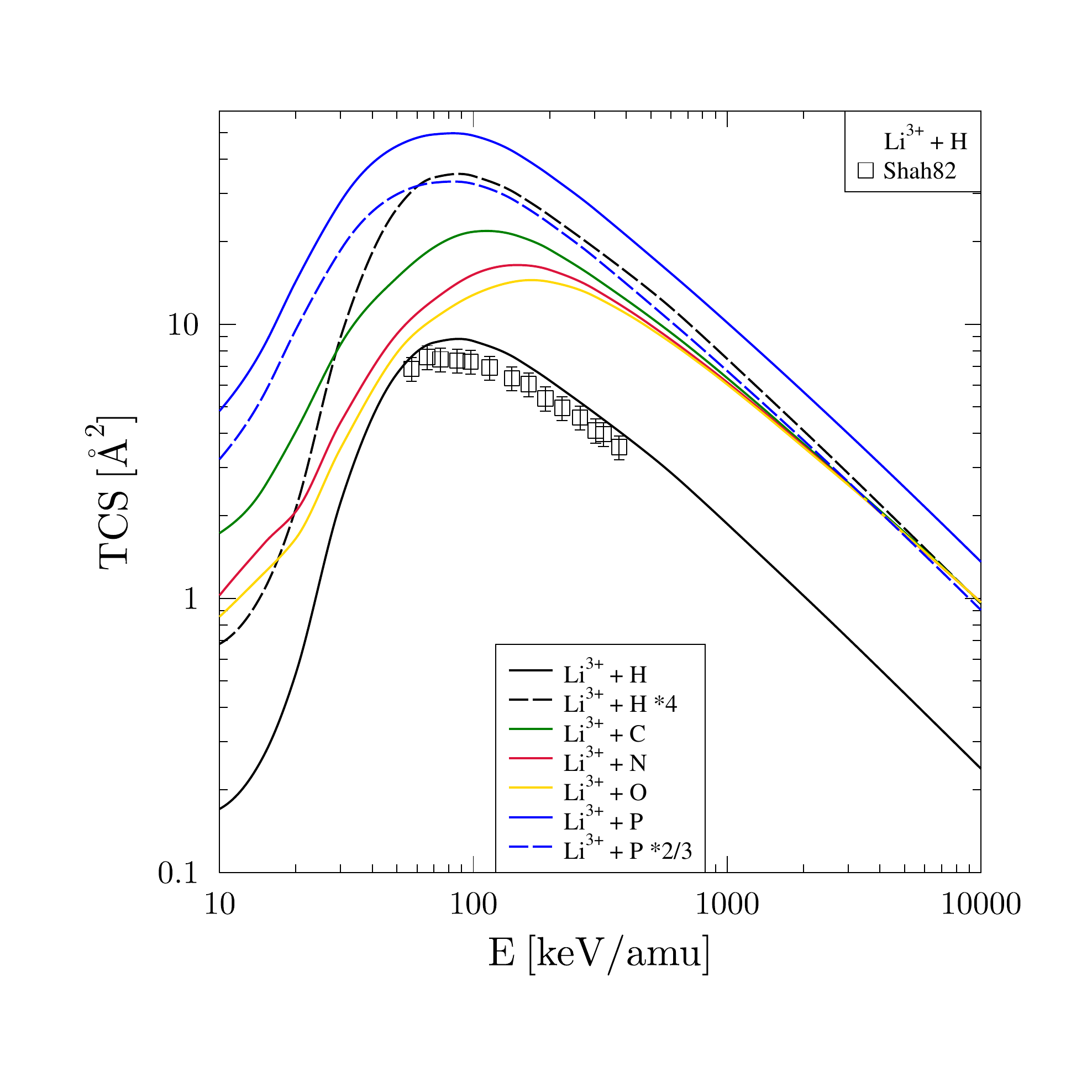}}

\end{array}$
%}
\caption{%
In the left panel the total (net) ionization cross sections for He${}^{2+}$ ions colliding with atoms H, C, N, O, P is shown (solid lines) as calculated with the TC-BGM
using basis sets described in the text,
while the right panel is for the case of   Li${}^{3+}$ projectiles. The dashed lines indicate how in the limit of high
$E$ the merging results for the heavier target atoms can be related to the values for atomic hydrogen by using a scale factor. The He${}^{2+} -  \rm H$ results are compared
with the convergent close coupling calculation of Ref.~\cite{Faulkner_2019}, and the experimental data are from Ref.~\cite{Shah_1981,Shah_1988} in this case,
and from Ref.~\cite{Shah_1982} for Li${}^{3+} -  \rm H$ collisions.
}
\label{fig:Abb18and19}
\end{center}
\end{figure}

Fig.~\ref{fig:Abb18and19} displays the net ionization cross sections that form the ingredients for direct application of the IAM-AR and IAM-PCM calculations described in the
next subsection, from which then a scaling model is derived and verified for ion-molecule collisions. The results demonstrate the non-perturbative nature of these calculations
which were obtained using an exchange-only density functional (optimized potential) method for the target atoms other than hydrogen. 

While a detailed comparison of 
charge-state correlated cross sections would benefit from dynamical screening, the net ionization cross sections are deemed reliable from such a frozen-potential approach
for proton impact. For $\rm C^{4+}-Ne$ collisions at $E=20 \ \rm keV/amu$ 
we found in Ref.~\cite{PhysRevA.64.012711} that the total ionization cross sections are reduced by $25 \%$ when response
is included for both the target and projectile potentials. For higher energies the effect is less pronounced. It can be debated whether IAM-PCM can be based
on atomic calculations with response, and our current approach is that we use frozen-potential calculations even for higher projectile charges.

The basic shapes of these cross sections roughly do not change with increasing projectile charge $Q$, and there are differences in shape for the various target atoms.
In the limit of high energies we observe the known scaling with $Q^2$, but in the vicinity of the maxima in the cross sections this scaling factor is reduced.
There is a small, but noticeable shift in the position of the maxima as $Q$ increases, and then at low collision energies ionization becomes less efficient with increasing
projectile charge so that the net ionization cross sections become comparable even though $Q$ is increased. At even lower energies one can understand
the inefficiency of ionization by multiply charged projectiles not only as the competition with capture channels, but also on the basis of (avoided) adiabatic energy 
level crossings. The TC-BGM calculations are deemed capable of describing such behavior reasonably well. A detailed comparison of shapes for $Q=2,3$ then
shows a steepening of the rise in the cross section starting from low $E$ values.

The ion-atom collision cross section calculations are challenging and require significant computing resources. Comparison with the experimental data for atomic hydrogen
is possible  (for $\rm He^{2+}$ and $\rm Li^{3+}$ projectiles
the data are from Ref.~\cite{Shah_1981,Shah_1988} and~\cite{Shah_1982}  respectively)  The comparison shows that converged theory is slightly higher than experiment.
For p-H collisions (cf. Fig.~6 in Ref.~\cite{hjl19}) a similar situation arises. 
The agreement of our calculation with the convergent close coupling approach (CCC) for $\rm He^{2+}$ impact~\cite{Faulkner_2019} is as good as can be expected from
two complementary large-scale approaches. 

The cross sections shown in Fig.~\ref{fig:Abb18and19} for $Q=2$ and $Q=3$ projectiles display a scaling behavior (apparent by the log-log scaling of the axes), and it is this
scaling which leads to scaling for molecular targets when treated by the additivity rule, i.e., the IAM-AR approach. This scaling behavior forms the starting point for the next section.

\subsection{Scaling model for ion-molecule collisions}
\label{sec:model2}

\begin{figure}
\begin{center}$
%\resizebox{0.6\textwidth}{!}{%
\begin{array}{ccc}\hspace{-1. truecm}
\resizebox{0.41\textwidth}{!}{\hspace{-1.5 truecm}\includegraphics{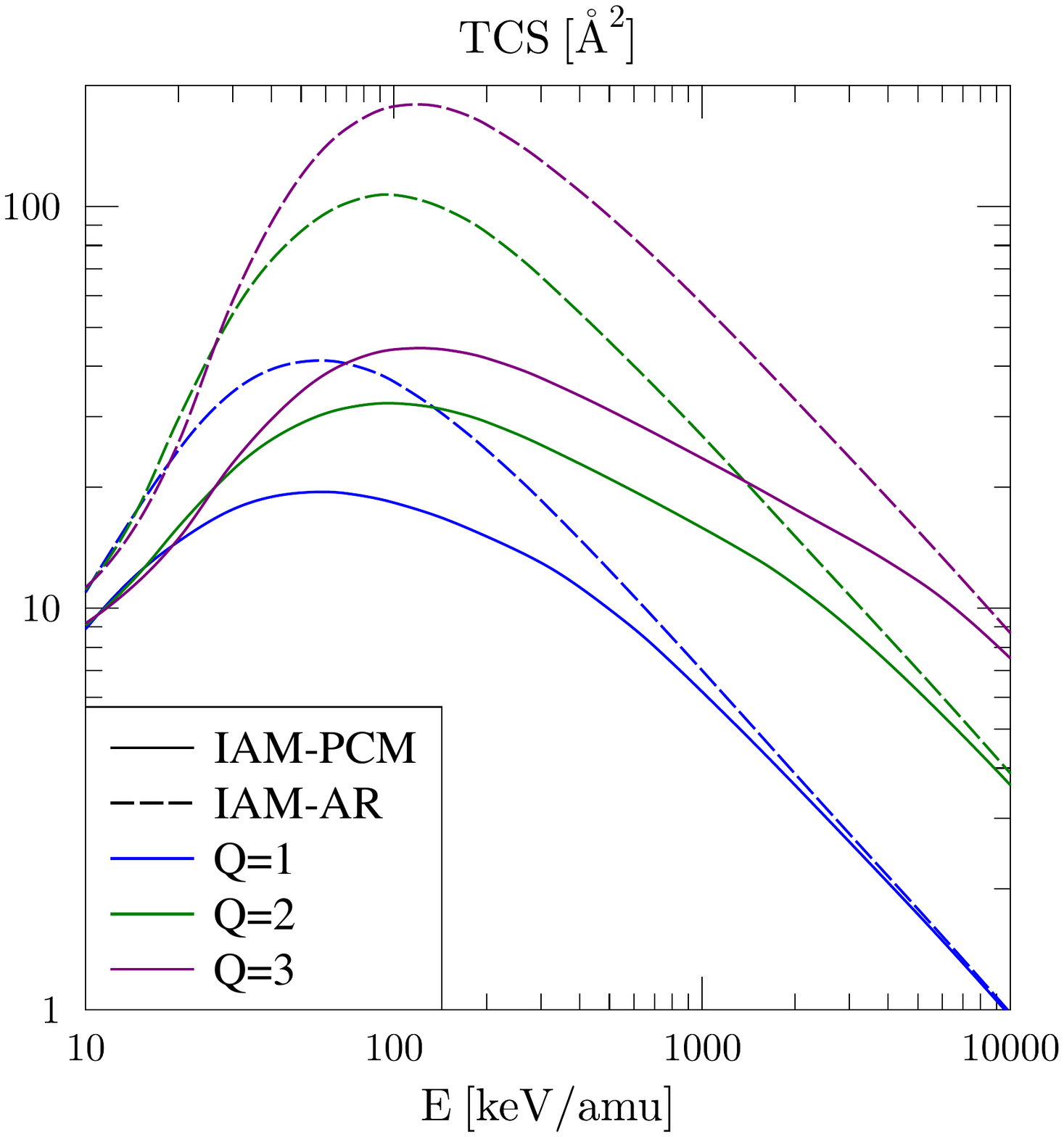}}\hspace{-1.5 truecm}&
\resizebox{0.41\textwidth}{!}{\hspace{-1.5 truecm}\includegraphics{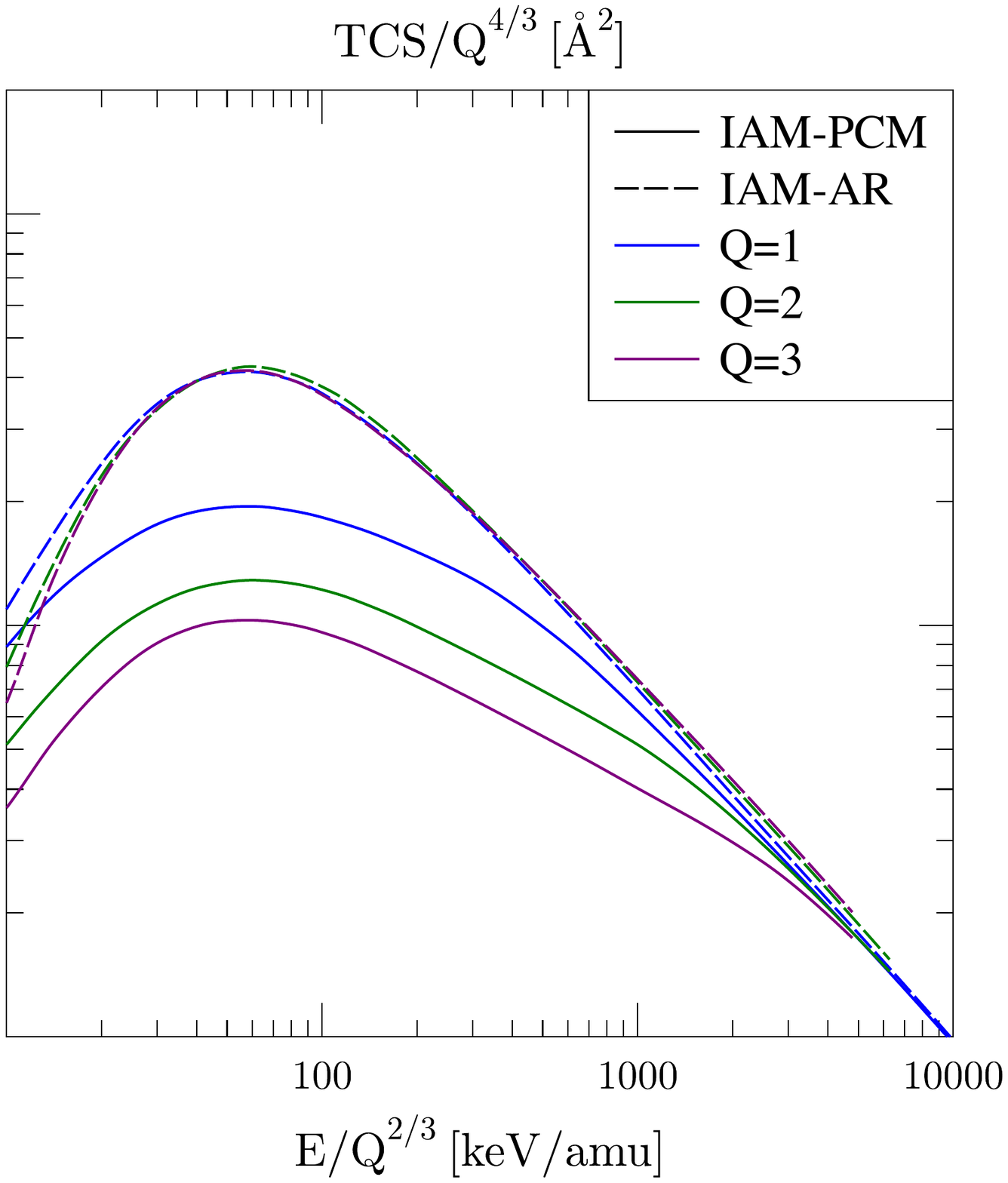}}\hspace{-1.5 truecm}&
\resizebox{0.41\textwidth}{!}{\hspace{-1.5 truecm}\includegraphics{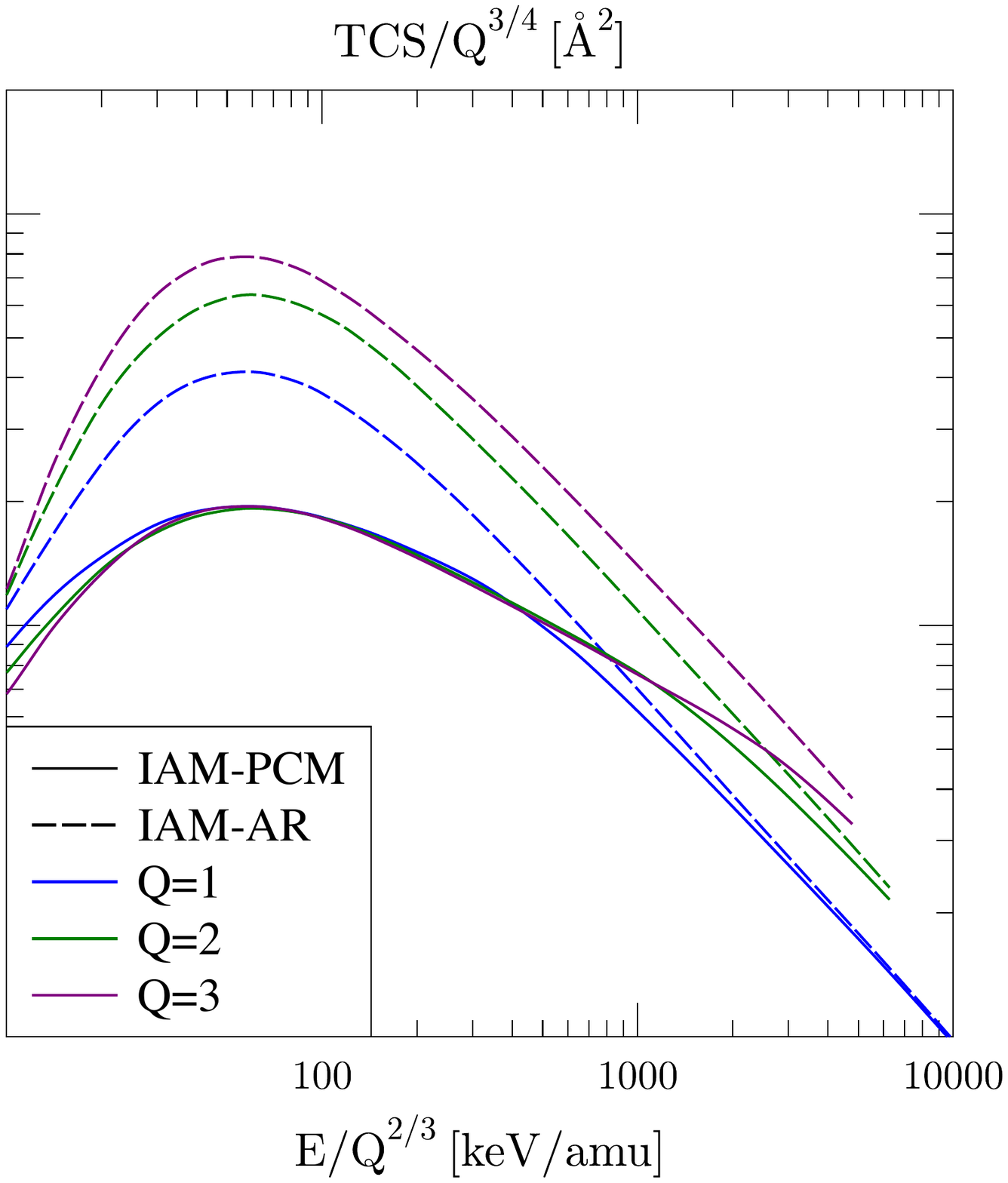}}

\end{array}$
 \caption{
   The left panel shows the total (net) ionization cross sections produced from uracil ($\rm C_4 H_4 N_2 O_2$) by charged-ion impact for $Q=1,2,3$ as calculated in the IAM-PCM (solid lines), and in the IAM-AR (dashed lines). In the left panel $Q=1$ appears at the bottom and $Q=3$ at the top.  The middle panel shows a scaled net ionization cross section for the IAM-AR model (dashed lines falling onto a single curve to within  5 \% for $E> 20 \ \rm keV/amu$), and the the IAM-PCM calculations which do not obey this scaling behavior in the energy range shown (here $Q=1$ appears at the top, while $Q=3$ is at the bottom). The right panel shows a scaled net ionization cross section that works for the IAM-PCM (solid lines agreeing to within 5 \% in a range enclosing the maximum, i.e., for $30 \le E \le 500 \ \rm keV/amu$), while the IAM-AR results show $Q=3$ at the top and $Q=1$ at the bottom.
   }
   \label{fig:Abb21to23}
\end{center}
\end{figure}

Fig.~\ref{fig:Abb21to23} forms the basis for the discussion. The empirical scaling models aim to capture the essence of the data in the non-perturbative regime where the
net ionization cross sections have their maxima. The left panel of Fig.~\ref{fig:Abb21to23} displays the net ionization cross sections for the two models, which have the following
characteristics: {\it (i)} with increasing projectile charge $Q=1,2,3$ the difference between IAM-PCM (solid lines) and IAM-AR (dashed lines) results grows dramatically. This is caused
by the increase in the atomic cross sections with projectile charge $Q$ (except at the lowest collision energies) and thus an increase in the overlap effect which the IAM-PCM takes into account; 
{\it (ii)} the position of the maxima in these cross sections changes very similarly with $Q$ for the two models, which implies that a common energy scaling can be applied, particularly
if one is interested in scaling the cross sections in the vicinity of the maximum.

For the IAM-AR (middle panel of Fig.~\ref{fig:Abb21to23}) the energy scaling for the maximum implies that it is useful to introduce
\begin{equation}
E' = E \ Q^{-2/3}
\label{eq:eq1}
\end{equation}
in order to find a common curve, and that the net cross sections should be scaled in accord with
\begin{equation}
\sigma^{\rm AR}_{Q=1} (E') = \sigma^{\rm AR}_Q(E) \ Q^{-4/3} 
\label{eq:eq2}
\end{equation}
in order to obtain a single universal curve for $\sigma^{\rm AR}_{Q=1} (E')$ where the only $Q$-dependence is through the scaled energy $E'$.
Here $\sigma^{\rm AR}_Q(E)$ is the IAM-AR net ionization cross section for a projectile with charge $Q$ and energy $E$ colliding with the molecule.

In the right panel of Fig.~\ref{fig:Abb21to23} the IAM-PCM data are presented while applying the same energy scaling. Since in the vicinity of the maximum the IAM-PCM cross
section grows differently with $Q$ than the IAM-AR result quite a different scaling factor is needed: $Q^{-0.75}$ vs $Q^{-4/3}$. 
We then arrive at
\begin{equation}
\sigma^{\rm PCM}_{Q=1} (E') = \sigma^{\rm PCM}_Q(E) \ Q^{-0.75} \ .
\label{eq:eq3}
\end{equation}

If the scaling works the reduced cross sections $\sigma^{\rm AR}_{Q=1} (E')$ and $\sigma^{\rm PCM}_{Q=1} (E')$ must correspond to
the calculated cross sections for proton impact at energy $E'$. The IAM-PCM and IAM-AR results merge in the limit of high $E'$, since the atomic
cross sections become small in this limit, such that the geometric overlap effect disappears. 
We note that others have made attempts at scaling ion-molecule cross sections using classical-trajectory
calculations~\cite{Otranto2019} or distorted-wave models~\cite{mendez2019ionization,alej2020universal}, 
and these attempts have so far not led to an understanding of scaling in the regime of the cross section maxima,
but have focused on the high-energy behavior.

The present calculations obey the following behavior in the limit of high energies: the IAM-PCM goes over into the IAM-AR in this limit because the cross sections
become small. Where exactly this occurs depends on the value of the projectile charge $Q$. The left panel of Fig.~\ref{fig:Abb21to23} indicates that for protons
this merge happens at about $E=2000 \ \rm keV/amu$, and at substantially higher energies for $Q>1$ to the point where for these higher projectile charges the
overlapping effect contained in the IAM-PCM cannot be ignored in the regime of interest (below 10 MeV/amu).

The scaling behavior of the IAM-AR is compatible with the behavior known for the ion-atom cross sections, i.e., the Bethe-Born limit which is obeyed
by the TC-BGM calculations~\cite{hjl18,hjl19}.
Since this limit applies to proton-atom collisions we can argue that for any molecular target the IAM-AR net ionization cross section 
for a projectile of charge $Q$ obeys 
\begin{equation}
\sigma^{\rm AR}_Q (E) \ \rightarrow \ Q^{4/3} \frac{A \ln{E' +B}}{E'} \ = \ Q^{4/3}\frac{A \ln{\left( E Q^{-2/3} \right)}+B}{E Q^{-2/3}}  \ = \ \frac{Q^2}{E}(A \ln{E} + B')\ ,
\label{eq:eq4}
\end{equation}
and therefore is proportional to the square of the projectile charge.

The right panel of Fig.~\ref{fig:Abb21to23} displays an interesting phenomenon: while the IAM-PCM results do go over into the corresponding IAM-AR ones at the highest
collision energies there is a cross-over phenomenon connected with the shape change. While the IAM-PCM result for protons merges with the IAM-AR result smoothly,
for the cases of $Q=2$ (green) and $Q=3$ (red) we find extended regions around $E=1 \ \rm MeV/amu$ with a characteristic slope that is not as steep as the
perturbative high-$E$ result. This will turn out to be of even bigger importance discussed below in the context of $Q=4-8$ projectiles at $E=1-6 \ \rm MeV/amu$.
At high energies the  IAM-PCM scaling expression (3) (which captures the behavior near the maximum in the cross section) is replaced by the IAM-AR scaling, which is
also obeyed by the IAM-PCM in this limit. A parametrization is
introduced to switch from the IAM-PCM form of scaling which works at intermediate energies $E$ to the IAM-AR scaling in order to construct a scaling formula for
the IAM-PCM results that is valid at all energies.

Now that we understand the scaling behavior of the IAM-PCM cross sections we proceed with developing a parametrization that will allow to predict cross sections for
collisions with higher projectile charges $Q$, for which the computation of ion-atom collisions using the TC-BGM becomes very challenging. The idea is to find a parametrization
that allows to derive $Q$-fold charged projectile collisions from molecules on the basis of proton-impact collisions (at scaled energies),
confirm its validity for $Q=2,3$ and then compare with experiment which is available for $Q=4..8$. 
We can expect such empirical scaling to work at energies $E>40 \ \rm keV/amu$ on the basis of the data shown in the right panel of Fig.~\ref{fig:Abb21to23}.
The scaling of the IAM-PCM cross section maxima with $Q^{0.75}$ is an empirical result for uracil. In general, we assume scaling with $Q^\beta$, with the
implication that $\beta=0.75$ for uracil and that it can have different values for other molecules where the cross sectional area overlap effects are different.

In order to accommodate the change in behavior which is apparent in the right panel of Fig.~\ref{fig:Abb21to23} we parametrize the representation of the 
IAM-PCM cross section by a switching function which keeps the IAM-PCM scaling in the vicinity of the maximum, but then with increasing collision energy shifts over
to the known IAM-AR scaling of the cross sections. This is not an attempt to modify the IAM-PCM result, however, it is just a convenient way to represent it making use of
both scaling regimes obeyed by the IAM-PCM calculations.

The switching function is defined as
\begin{equation}
a(E) = \left[ 1+ \left( \frac{E/E_0}{Q-1}\right)^{\frac{4}{\sqrt{Q}}} \right]^{-1} \ ,
\label{eq:eq5}
\end{equation}
with the switching point chosen as $E_0=2000 \ \rm keV/amu$ and $Q \ge 2$. 
Note that the case of anti-proton impact ($Q=-1$) is explicitly excluded. While in the high-$E$ limit ionization by protons and anti-protons does yield the
same amount of net ionization, the physics is very different in the vicinity of the cross section maximum 
(saddle-mechanism for positively charged projectiles, pushing charge out of the way for anti-proton impact).
The parametrization of the IAM-PCM net ionization cross section scaling now takes the form
\begin{equation}
\sigma^{\rm PCM}_Q(E) = a(E) Q^\beta \sigma^{\rm PCM}_{Q=1}(E') + (1-a(E)) Q^{4/3} \sigma^{\rm AR}_{Q=1}(E') \ .
\label{eq:eq6}
\end{equation}
This parametrization of the IAM-PCM cross sections for $Q$-fold charged ion impact in terms of the proton-impact cross sections evaluated at $E'$ depends
on a single parameter $\beta$ whose value is connected with the `density' of independent atoms, and which needs to be determined by tracking
the movement of the maxima with charge in the IAM-PCM cross sections. A discussion of detailed results for the uracil target (for which $\beta=0.75$ was shown to work) 
follows, and a table of values of $\beta$ for other molecules is given further below.

We emphasize that the role of the interpolation scheme (6) is to provide the scaling description of the IAM-PCM results. The appearance of the IAM-AR cross section in (6) 
follows from the construction and is natural (IAM-PCM goes over into IAM-AR in this perturbative limit). We also note that the choice of interpolating form (5) for $a(E)$ is not
unique, but that we can demonstrate that it works at the few-percent level of accuracy.

\begin{figure}
\begin{center}
%\resizebox{0.6\textwidth}{!}{%
\resizebox{0.6\textwidth}{!}{\includegraphics{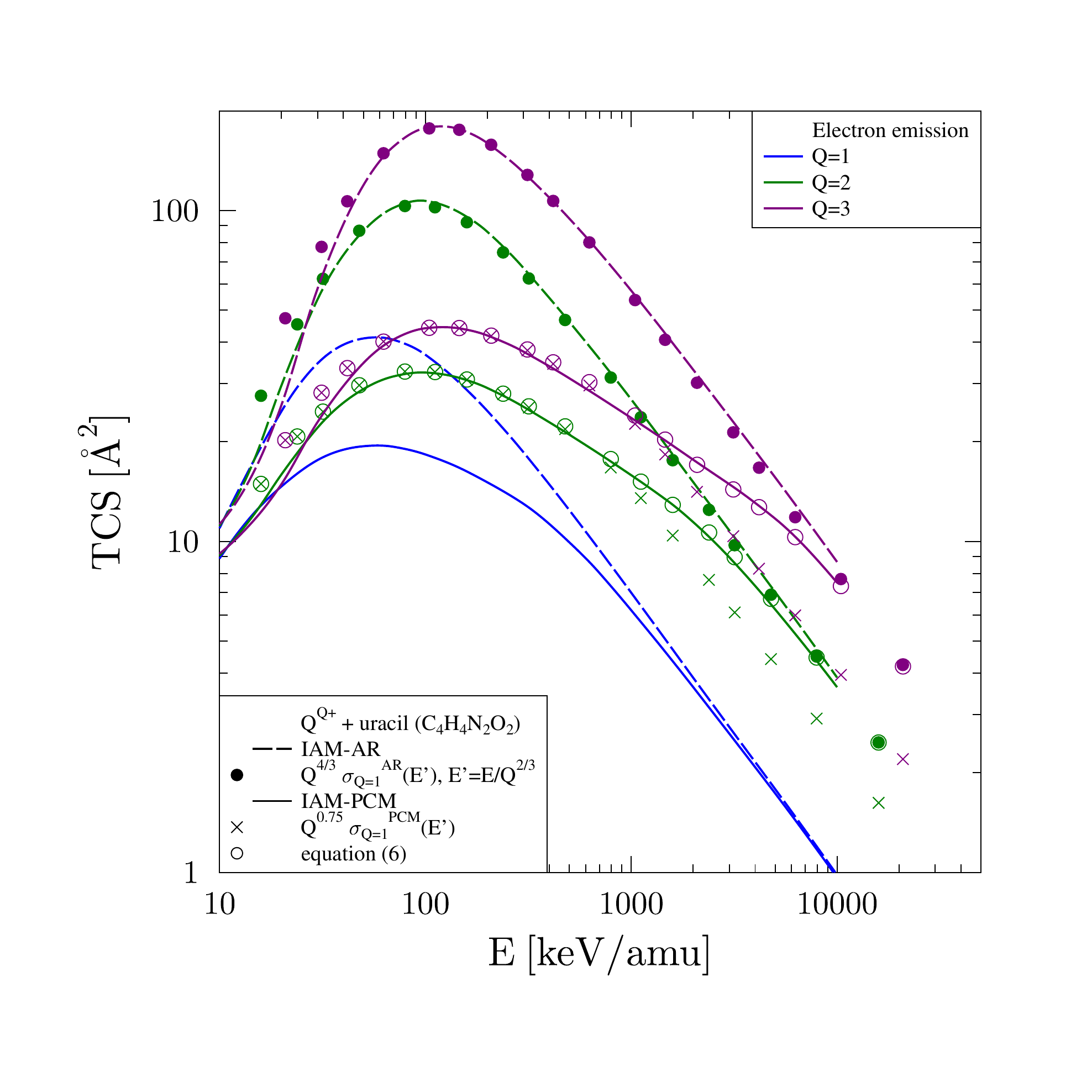}}  

\caption{
   Net ionization cross sections for projectile charges $Q$ impinging on uracil. Solid and dashed lines are the IAM-PCM and IAM-AR calculations respectively for $Q=1,2,3$.
   Solid circles: IAM-AR scaling prediction based on $Q=1$ IAM-AR calculations. Crosses: IAM-PCM scaling predictions based on $Q=1$ IAM-PCM calculations which track the respective maximum in the cross section but fail at high energies $E$. 
   Open circles: Parametrization results for the IAM-PCM cross sections which switch from IAM-PCM scaling for the maximum to IAM-AR scaling for the high-$E$ regime.
   The radius of the open circles corresponds to a deviation of $5 \%$ in the cross section.
      }
   \label{fig:Abb24}
\end{center}
\end{figure}

In Fig.~\ref{fig:Abb24} the effectiveness of the parametrization for energies $E>40 \ \rm keV/amu$ is demonstrated. The IAM-AR results are reproduced by the
simple scaling law, as suggested by data shown in the middle panel of Fig.~\ref{fig:Abb21to23}. The IAM-PCM scaling results (crosses) track the data well
around the maxima in the cross sections, but show some weakness at low energies ($E<40 \ \rm keV/amu$) for reasons discussed before. Ionization
by highly charged ions becomes ineffective at low energies as one moves towards the quasi-molecular regime and due to increasing competition 
from capture channels. 

Given that the IAM-PCM cross section for proton impact converges to the IAM-AR cross section at energies of scale $E_0  = 2000 \ \rm keV/amu$ it is possible
to simplify the parametrization and eliminate the IAM-AR cross section $\sigma^{\rm AR}_{Q=1}$ while admitting a small error. This is due to the fact
that the switching function~(\ref{eq:eq5}) deviates significantly from unity only at very high collision energies $E > E_0$.
This simplified parametrization
then becomes
\begin{equation}
\sigma^{\rm PCM}_Q(E) = \left [ a(E) Q^\beta  + (1-a(E)) Q^{4/3} \right ] \sigma^{\rm PCM}_{Q=1}(E')  \ .
\label{eq:eq7}
\end{equation}
This form allows one to take experimental data for $\sigma^{\rm net}_{Q}(E)$ and relate it directly to the proton impact result $\sigma^{\rm net}_{Q=1}(E')$
in order to test whether the scaling prediction from the IAM-PCM theoretical model translates into measured physical evidence. The only required additional piece of
information for a given molecule is the $\beta$ parameter which has to be found on the basis of IAM-PCM calculations for $Q=1,2,3$. 
A list of parameter values is given in Table~\ref{table:beta}
for a few characteristic examples. The value of $\beta$ is larger for simple molecules such as water or methane, and decreases when the number of constituent atoms increases.

\begin{table}[ht]
\caption{parametrization model parameter $\beta$  for the investigated molecules.}
\centering
\begin{tabular}{c c c c}
\hline\hline
formula & name & group & $\beta$ \\ [0.5ex] % inserts table %heading
\hline
$\rm H_2O$&water&&1.1 \\
$\rm CH_4$&methane&alkane&1.05 \\
$\rm C_4H_4N_2O_2$&uracil&pyrimidine &0.75 \\
$\rm C_5H_5N_5$ &adenine & purine & 0.7 \\
$\rm C_5H_{11}N O_2$ & L-valine & amino acid & 0.65 \\
$\rm C_{10}H_{14}N_5 O_6 P$ & dAMP & nucleotide & 0.6\\ [1ex]
\hline
\end{tabular}
\label{table:beta}
\end{table}

\begin{figure}
\begin{center}
%\resizebox{0.6\textwidth}{!}{%
\resizebox{0.6\textwidth}{!}{\includegraphics{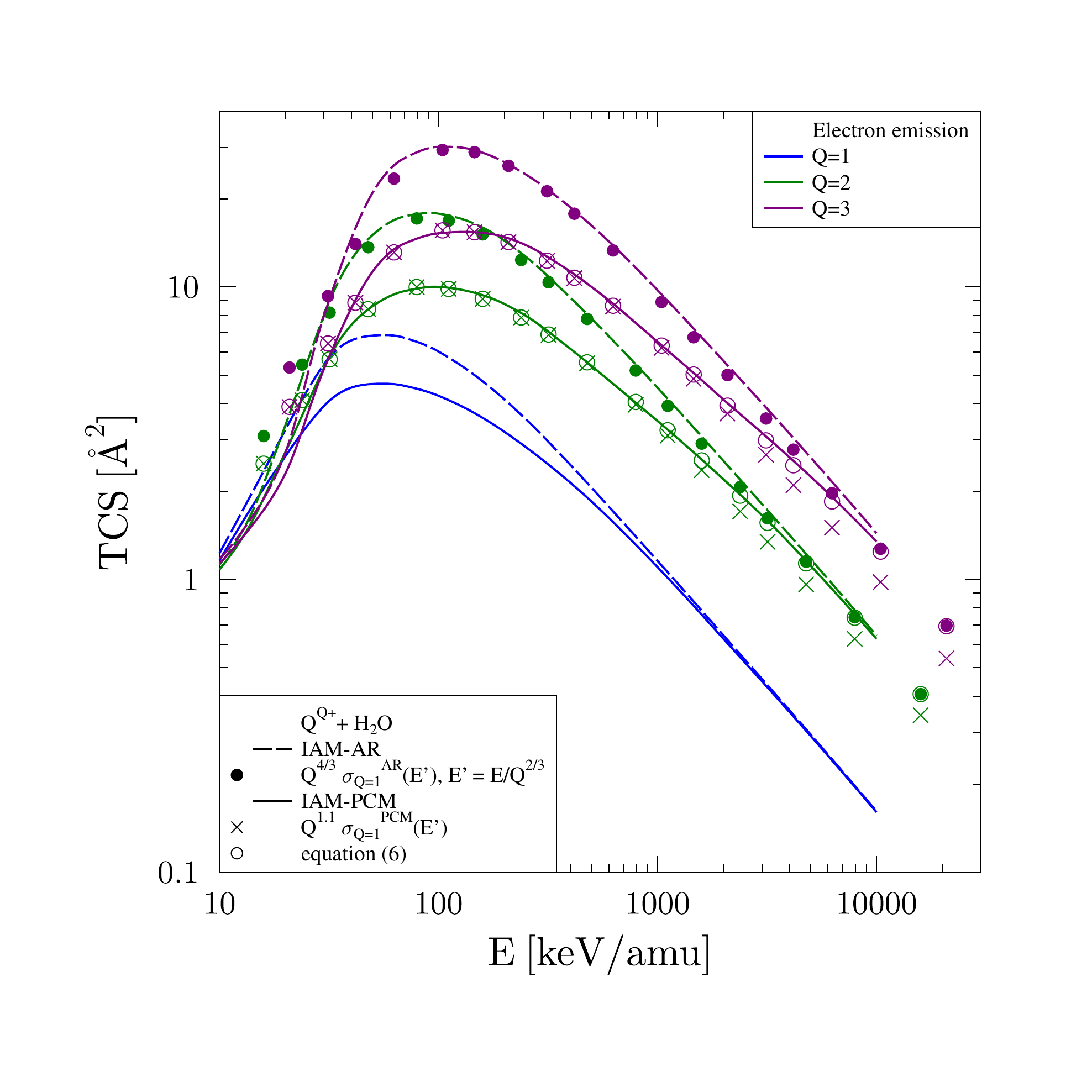}}  
   
\caption{Same as in Fig.~\ref{fig:Abb24} but for the water molecule  target. 
}
 \label{fig:Abb25}
\end{center}
\end{figure}

We conclude this section by a demonstration of the effectiveness of the parametrization for the water molecule target in Fig.~\ref{fig:Abb25} for which the
$\beta$ parameter is larger than for uracil. We begin by noting that the $\rm p-H_2O$ IAM-PCM cross sections have been corrected compared to our previous
work~\cite{hjl16,hjl18}  in that the correct bond length has been implemented here. We have also tested the IAM-PCM calculated cross sections against an exact 
geometric calculation of the overlapping circular areas which is possible for a triatomic molecule, and found that the IAM-PCM implementation matches
the geometric calculation to excellent precision.

As a consequence of the corrected bond length error the IAM-PCM net ionization cross section now is further reduced around the maximum as compared to the IAM-AR
calculation, and the IAM-PCM and IAM-AR results now merge only at an energy of $E=1000 \ \rm keV/amu$. We also verified that $\rm H_2O$ was the only target
molecule for which a bond length error happened (it was an angstrom vs atomic unit problem, and the previously made error is apparent in Figs. 2a and 3a
of Ref.~\cite{hjl18}).

Concerning the comparison of scaling behavior for uracil $(\beta=0.75)$ and water $(\beta=1.1)$, i.e., comparing Fig.~\ref{fig:Abb24} to Fig.~\ref{fig:Abb25} 
we can provide the following comments. In the case of water the larger $\beta$-parameter value implies that the feature associated with the maximum in
the net proton-molecule collision (for which IAM-PCM, i.e., atomic cross sectional area overlap is giving the strongest effect) acquires a higher scaling with $Q$.
The progression of cross section values at maximum is (in units of \AA${}^2$) $\{4.6, 10, 15\}$ for $Q=\{1,2,3\}$. For the uracil target the corresponding progression is 
$\{19, 32, 44\}$. The decrease of the $\beta$ value with the complexity of the molecule is connected with the fact that
overlap effects become more important with the growing number of atoms and the IAM-PCM takes this into account.

\section{Comparison with experiments}
\label{sec:expt}

\subsection{Collisions with uracil ($\rm C_4 H_4 N_2 O_2$)}
\label{sec:expt1}

Total ionization cross section measurements for uracil targets have a bit of a history. Net ionization can be obtained by collecting all electrons produced
during the collision, a measurement process know from ion-atom collisions~\cite{Rudd85a} (called $\sigma_-$), but that is not how all the data were obtained. 
Time-of-flight mass spectroscopy was used to determine the fragment yields, and some analysis went into finding the fragment distributions following
electron capture versus direct ionization~\cite{PhysRevA.81.012711} for proton impact at velocities that match those of the valence electrons (20-150 keV/amu).
Doubly charged uracil is not observed, so it is likely to fragment before the mass analysis as observed by coincidence spectroscopy of fragments~\cite{PhysRevLett.107.023202}. 
This must also be the case for higher-projectile charge impact, such as 
$\rm C^{4+}$ and $\rm O^{6+}$~\cite{PhysRevA.85.032711,Agnihotri_2013}, where doubly-charged uracil is absent from the mass spectra. 
Tabet et al.~\cite{PhysRevA.82.022703} found very large values for the direct ionization and capture cross sections in this energy range, for $E=80 \ \rm keV$
proton impact direct ionization is reported to reach $177 \pm 35$  \AA ${}^2$. Itoh et al.~\cite{Itoh13} determined absolute differential cross sections
at higher energies by detecting the electron yield and by comparing with calculations, they demonstrated that the high cross section value of Ref.\cite{PhysRevA.82.022703}  was in conflict with their findings for $\sigma_-$. 

The heavy ion measurements in the same velocity regime~\cite{PhysRevA.85.032711,Agnihotri_2013} were normalized by using features in the differential electron
emission: KLL Auger lines from the constituent atoms C, N, O were observed. Absolute normalization at a high energy (3.5 MeV/amu) was obtained with an uncertainty
estimated at $\pm 20 \%$, and relative uncertainties for the data were assessed at  $\pm 12 \%$. The total cross sections were obtained by measuring the fragment
yields with some uncertainty coming from $\rm N_2^+$ contributions from collisions with the background gas. Unlike the proton impact measurements of Ref.~\cite{PhysRevA.82.022703},
however the projectile charge state after the collision was not determined. To ensure a proper separation from ionization and fragment production by pure capture
processes coincidence with an ejected electron was imposed. This procedure includes electron production from transfer ionization processes, but if one then
adds up all charged fragments one overestimates the contribution from transfer ionization to net ionization
(by a factor of two for the process where one electron is captured,
and one transferred to the continuum, and by more for higher-order capture plus higher-order ionization). At the higher velocities ($v=7-15$ Bohr units) this problem
does not exist, since capture plays a small role (except for capture from innermost shells, but these are small contributions).

\begin{figure}
\begin{center}$
%\resizebox{0.6\textwidth}{!}{%
\begin{array}{cc}
\resizebox{0.5\textwidth}{!}{\includegraphics{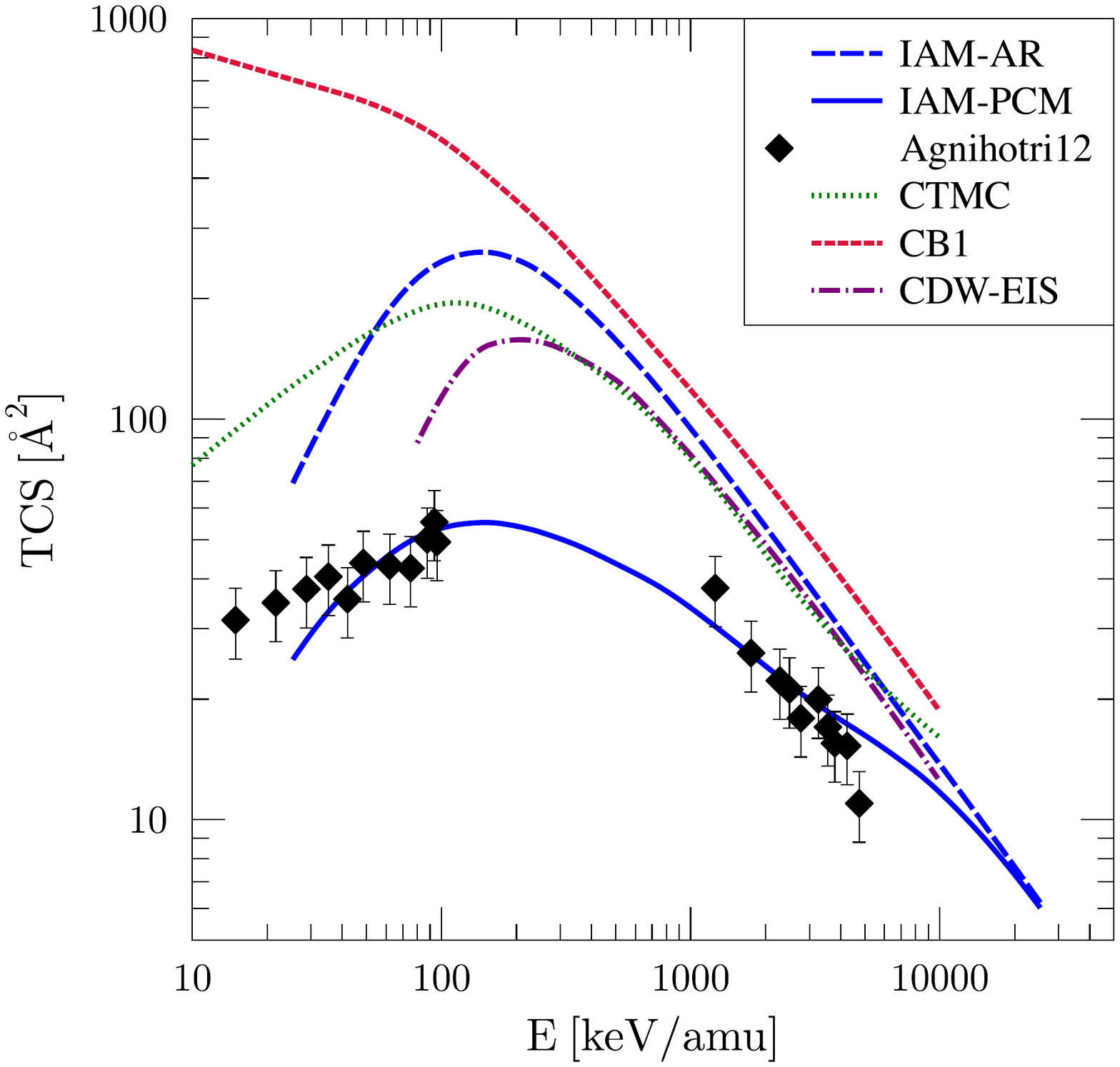}}&
\resizebox{0.5\textwidth}{!}{\includegraphics{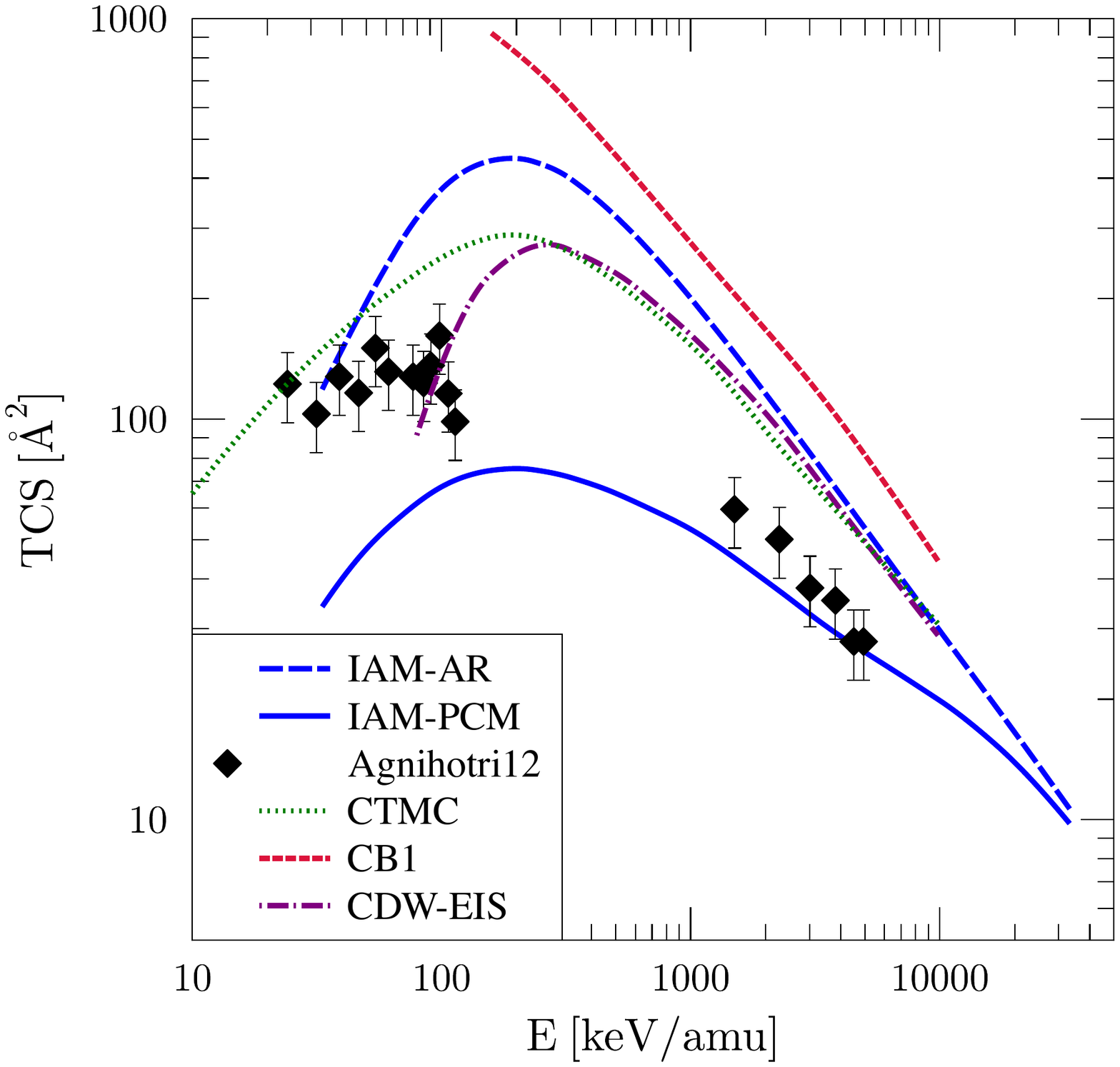}}

\end{array}$
   
\caption{
Net ionization cross sections for projectiles with charges $Q=4$ (left panel) and $Q=6$ (right panel) impinging on uracil ($\rm C_4 H_4 N_2 O_2$). Solid and dashed blue lines are the IAM-PCM and IAM-AR calculations respectively obtained by scaling the proton impact cross sections. The experimental data (diamonds with error bars) are for $\rm C^{4+}$ projectiles (left panel) and for 
$\rm O^{6+}$ projectiles (right panel)~\cite{PhysRevA.85.032711}. Also shown (dotted green line) are theoretical results from the classical trajectory method (CTMC)~\cite{Sarkadi_2016}, the first-order Born method with corrected boundary conditions (CB1) shown as a short-dashed red line, which is the highest curve), 
and from distorted wave theory CDW-EIS 
as a purple dash-dotted line (the latter two from Ref.~\cite{PhysRevA.85.032711}).
}
   \label{fig:Abb26and27}
\end{center}
\end{figure}

We now confront the predictions from the IAM-PCM scaling model which was established on the basis of $Q=2,3$ calculations with experimental and theoretical
data for higher projectile charges $Q=4,6$ for which experimental data exist both below the predicted maximum and above~\cite{PhysRevA.85.032711}.
The comparison provided in Fig.~\ref{fig:Abb26and27} reveals the following: at the high-energy end, i.e., for $E> \ \rm 1000 keV/amu$
the scaled IAM-PCM cross section does agree with the experimental data for both projectile charges
almost within error bars, but shows a somewhat different energy dependence. The energy dependence of the experimental data
follows the trend given by CTMC, CB1 (first Born model with correct boundary condition), and CDW-EIS in this regime, 
although these theories yield higher cross section values.
We note that both projectiles do have fully populated 
K-shells, i.e., we do not expect major anomalies in the cross sections from Auger processes.

At the lower end of collision energies, below $E=100 \ \rm keV/amu$ we find the comparison to be more complicated. For the case of $Q=4$
the present results join the two regimes of experimental data reasonably well, but fall short at the lower end where the observed scaling
behavior becomes less accurate as discussed in Section~\ref{sec:model2}. All other theories than scaled IAM-PCM provide estimates higher by at least a factor of four,
with CDW-EIS doing better, but showing an energy dependence that differs from the experimental and from the scaled IAM-PCM data trends.
Clearly, the situation calls for experiments to map out the location of the maximum in the net ionization cross section.

The comparison with the case of $Q=6$ seems to warrant a different conclusion: the experimental data are consistently higher than the scaled IAM-PCM
results by a factor of two in the regime to the left of the maximum in the IAM-PCM data. The experimental data trend is supported by the CTMC calculation,
which however failed in the case of $Q=4$. Again, we call for experiments in the regime that connects the low-$E$ and high-$E$ data shown in 
Fig.~\ref{fig:Abb26and27}. As explained at the beginning of the subsection the low-energy data in Ref.~\cite{PhysRevA.85.032711} 
may be affected by over-counting of transfer ionization contributions, and this may be a bigger problem for the system with the higher projectile charge,
but we have no further support from calculations to make this case. Therefore, the scaling of the net ionization cross section with $Q$ for impact velocities
that match the valence shell orbital speeds remains a mystery. 

Another non-perturbative calculation that could be trusted in this regime, the CTMC 
calculation of Ref.~\cite{Sarkadi_2016} matches the $Q=6$ experimental data (right panel) at low energies, then joins the CDW-EIS result which is a factor of two
above the high-energy data. 
For $Q=4$ (left panel) this calculation overestimates the data over the entire energy range, and still matches the CDW-EIS calculation at high $E$.
The comparison with the CTMC calculation is interesting, since (as explained in Ref.~\cite{PhysRevA.92.062704}) the calculation makes use of an effective multi-center potential
that matches quantum chemistry calculations, uses molecular orbital energies, and performs random molecular orientation sampling.

\begin{figure}
\begin{center}
%\resizebox{0.6\textwidth}{!}{%
\resizebox{0.6\textwidth}{!}{\includegraphics{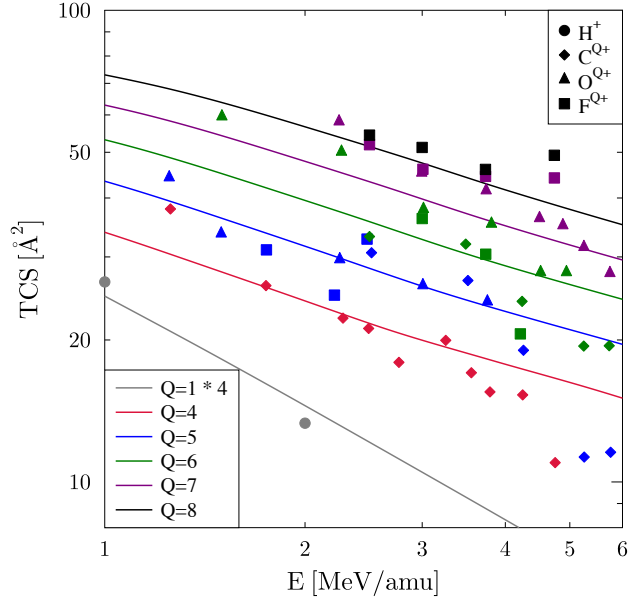}}  
   
\caption{
Experimental net (total) ionization cross sections for uracil ($\rm C_4 H_4 N_2 O_2$) bombarded by projectiles with charges $Q=1$ (Ref.~\cite{Itoh13}) and $Q=4-8$ 
   (Ref.~\cite{Agnihotri_2013, PhysRevA.85.032711}) are compared to the present IAM-PCM results. 
   The ordering of results is such that higher projectile charges $Q$ result in higher cross sections.
   The proton impact data are multiplied by a factor of four
   for comparison purposes and are shown in grey. The highly charged projectiles are distinguished by color (red, blue, green, magenta, black) for $Q=4-8$ respectively, and
   the projectile ions $\rm C, O, F$ are represented by diamonds, triangles and squares respectively. For clarity the experimental data are shown without uncertainties
   which are estimated to be $\pm 25 \%$ for the $Q \ge 4$ data.
  }
   \label{fig:Urazil_HE_Qabh}
\end{center}
\end{figure}

In Fig.~\ref{fig:Urazil_HE_Qabh} we present a  comparison with many available experimental ion-uracil ionization cross sections at high impact energies in order to explore
how the data scale. The most striking feature is the difference in the energy dependence between the case of $Q=1$ and the higher projectile charges
in the IAM-PCM. While the proton impact data are described well by both our results and the CDW-EIS model~\cite{mendez2019ionization}, the latter retains the energy dependence
for the higher charges $Q$. 
The difference in the energy dependence between $Q=1$ and the higher projectile charges for IAM-PCM is a consequence of the overlap effect when
the effective cross sectional area is computed: with increasing cross sections the areas representing the individual atoms increase, thereby causing
more overlap. Thus, we observe that while in the proton impact case the dependence still follows somewhat the proton-atom cross section energy
dependence,  a turnover happens between $Q=1$ and $Q=4$ in the IAM-PCM results. Note that all $Q>1$ cross sections
can be obtained simply from Eq.~(\ref{eq:eq7}) based on the input for IAM-PCM which are given in Table~A1 of Ref.~\cite{hjl19}. The slightly more accurate
data based on Eq.~(\ref{eq:eq6}) require the IAM-AR cross sections which can be assembled from ion-atom cross sections (there is no orientation 
dependence in this IAM, very much like in the CDW-EIS based calculations used, e.g., in Ref.~\cite{mendez2019ionization,alej2020universal}). Proton-atom cross sections
that are input to our IAM-AR result can be read off Fig.~6 in Ref.~\cite{hjl19}.

How well do the IAM-PCM results compare to the experimental data? Given that absolute normalization of ionization cross sections is a challenging 
task the  error estimates of the data ($\pm 25 \%$) are dominated by the normalization procedure. For some of the data the slope of the IAM-PCM results
matches experiment, particularly for $\rm O^{5+}$ and  $\rm C^{4+}$ projectiles over a wide range of energies. At the highest energies
measured the experimental data are sharply lower which does not help with the assessment as to whether the data follow the present IAM-PCM trend or not.
%For the higher charge states there is less clarity, and one question one may ask is whether the IAM-PCM needs to be computed on the basis of ion-atom
%cross sections where dynamical response in the target was taken into account. Such a mean field approach may reduce the net cross section, which in turn
%can lead to less overlap.
Concerning the $Q$ dependence
of the magnitude of the cross sections we observe that IAM-PCM matches the experimental data reasonably well, while the CDW-EIS and CTMC models show 
serious discrepancies, which is highlighted in greater detail in Fig.~\ref{fig:Fig8} below.

% One should also keep in mind that Auger processes may contribute to the experimental data, and the rise in the experimental
%$\rm F^{7+}$ data at high energies might be such an anomaly when compared to the trend in the $\rm O^{7+}$ data. It is unclear at present 
%to what extent Auger mechanisms
%can affect collision data with complex molecules, where close encounters with heavier atoms may be less important than in ion-atom collisions.

\begin{figure}
\begin{center}
%\resizebox{0.6\textwidth}{!}{%
\resizebox{0.6\textwidth}{!}{\includegraphics{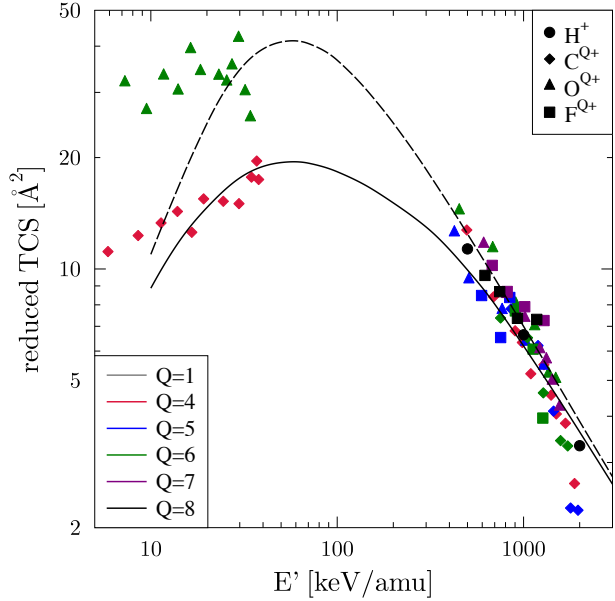}}

\caption{
The experimental net (total) ionization cross sections for uracil ($\rm C_4 H_4 N_2 O_2$) bombarded by projectiles with charges  $Q=4-8$ 
   (Ref.~\cite{Agnihotri_2013, PhysRevA.85.032711}) are turned into a reduced cross section $\sigma_{Q=1}$ using the simplified
   IAM-PCM scaling formula~(\ref{eq:eq7}) and compared to the IAM-PCM (solid line) and IAM-AR (dashed line) proton-uracil calculations.
   The three experimental proton impact data points from Ref.~\cite{Itoh13} are also shown (black solid circles). The symbols follow the same patterns
   as used  in Fig.~\ref{fig:Urazil_HE_Qabh}.
  }
   \label{fig:UrazilReduced}
\end{center}
\end{figure}

We now use the scaling behavior found on the basis of explicit IAM-PCM calculations for $Q=1,2,3$ (with some confirmation
from $Q=6$, which is not shown) to extract a reduced $Q=1$ cross section from the experimental
data for $Q=4-8$, and compare the result with the IAM-PCM $Q=1$ cross section, and show the IAM-AR $Q=1$ result for reference, as well.
Fig~\ref{fig:UrazilReduced} reveals a few rather interesting facts: the proton impact data at $E= \{0.5, 1.0, 2.0 \} \ \rm MeV$ span
the entire range of high-energy data from Ref.~\cite{PhysRevA.85.032711}, namely 1 to 6 MeV/amu.
A good part of the high-energy $Q=4-8$ experimental data is mapped into a band more or less between the IAM-PCM and IAM-AR
$Q=1$ cross sections, but the highest points near $E'=2 \ \rm MeV/amu$ (from $\rm C^{4+}$ and $\rm C^{5+}$) fall below. 
Data which would cover the $E'=50-400 \ \rm keV/amu$ would be most welcome in order to settle the question
whether the IAM-PCM prediction (solid line, which appears to work for $\rm C^{4+}$ projectiles, but not for the case of $\rm O^{6+}$)
is indeed correct. Proton impact data below $E=500 \ \rm keV/amu$ would also be most welcome.

\begin{figure}
\begin{center}$
%\resizebox{0.6\textwidth}{!}{%
\begin{array}{ccc}%\hspace{-1 truecm}
\resizebox{0.4\textwidth}{!}{\hspace{-1 truecm}\includegraphics{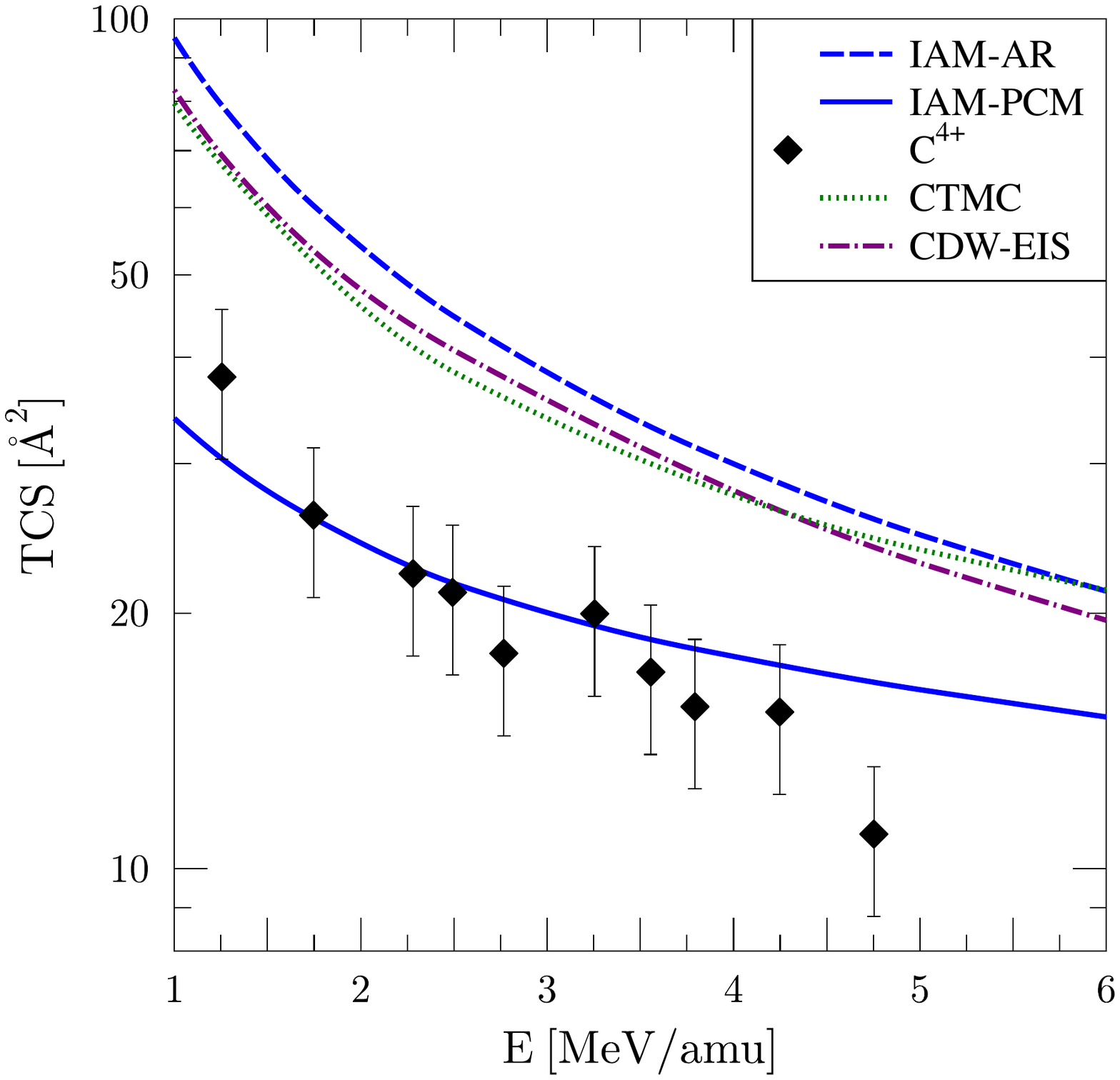}}\hspace{-1 truecm}&
\resizebox{0.4\textwidth}{!}{\hspace{-1 truecm}\includegraphics{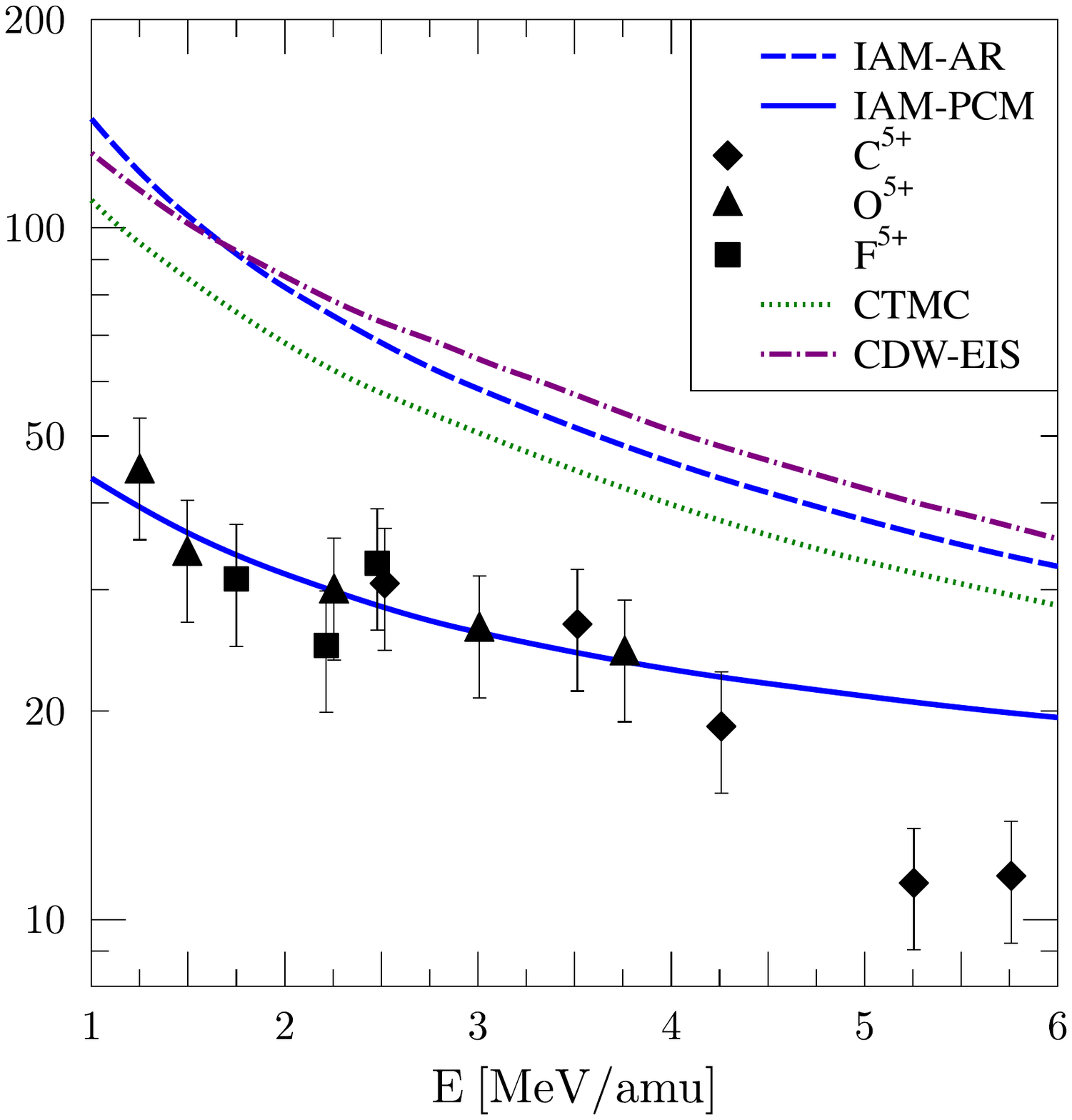}}\hspace{-1 truecm}&
\resizebox{0.4\textwidth}{!}{\hspace{-1 truecm}\includegraphics{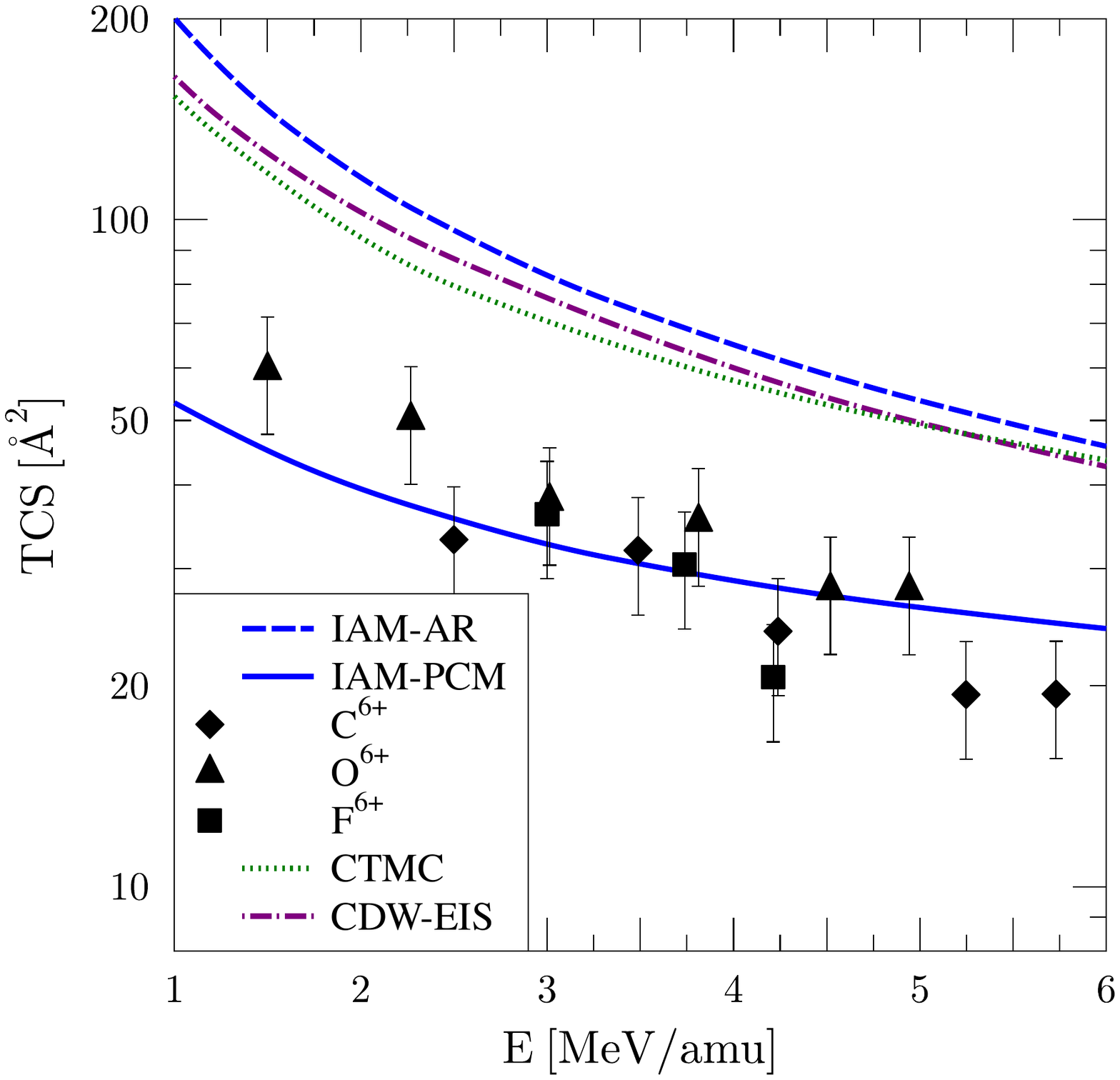}}\hspace{-1 truecm}\\[-80 pt]
\resizebox{0.4\textwidth}{!}{\hspace{-1 truecm}\includegraphics{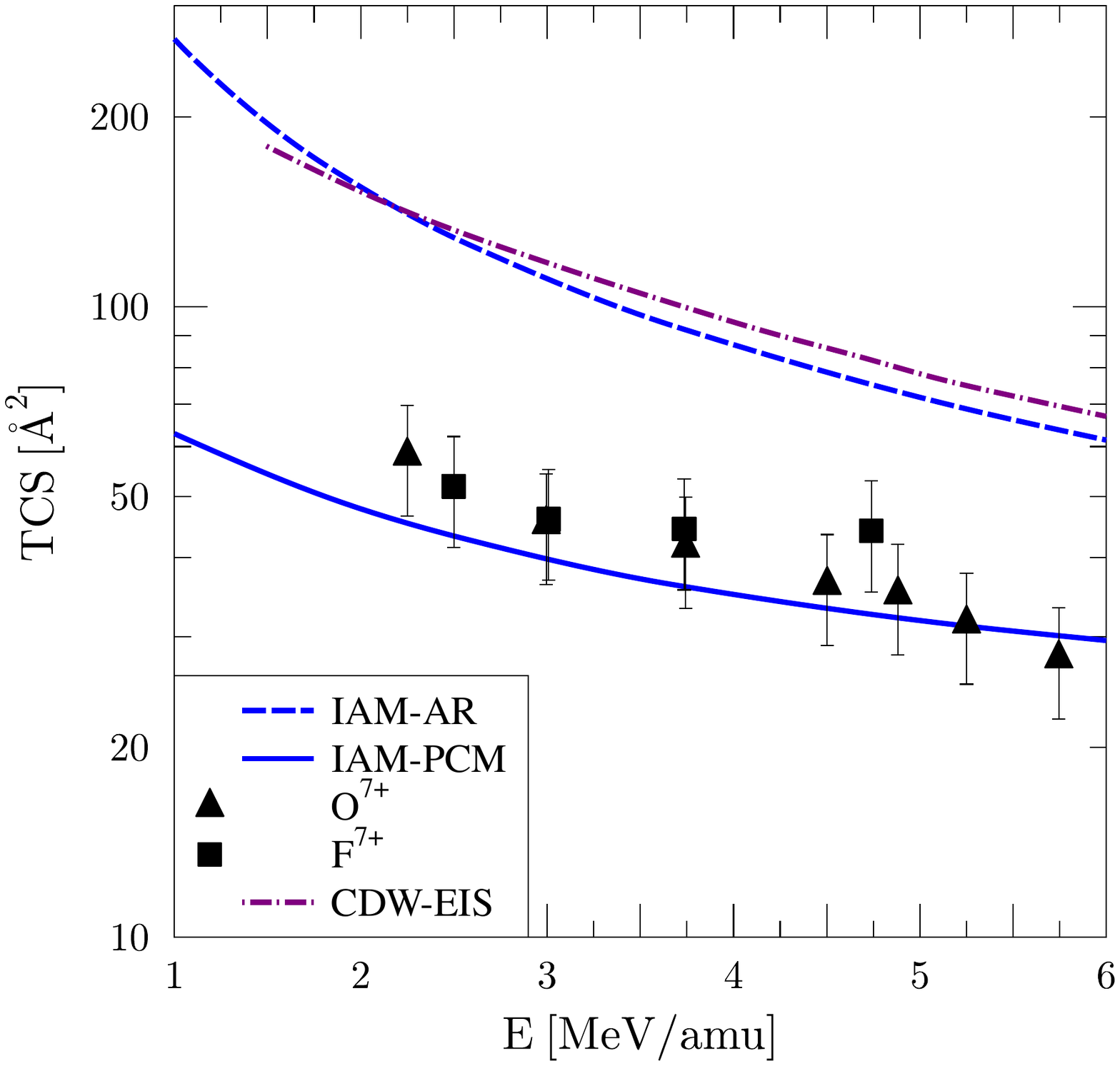}}\hspace{-1 truecm}&
\resizebox{0.4\textwidth}{!}{\hspace{-1 truecm}\includegraphics{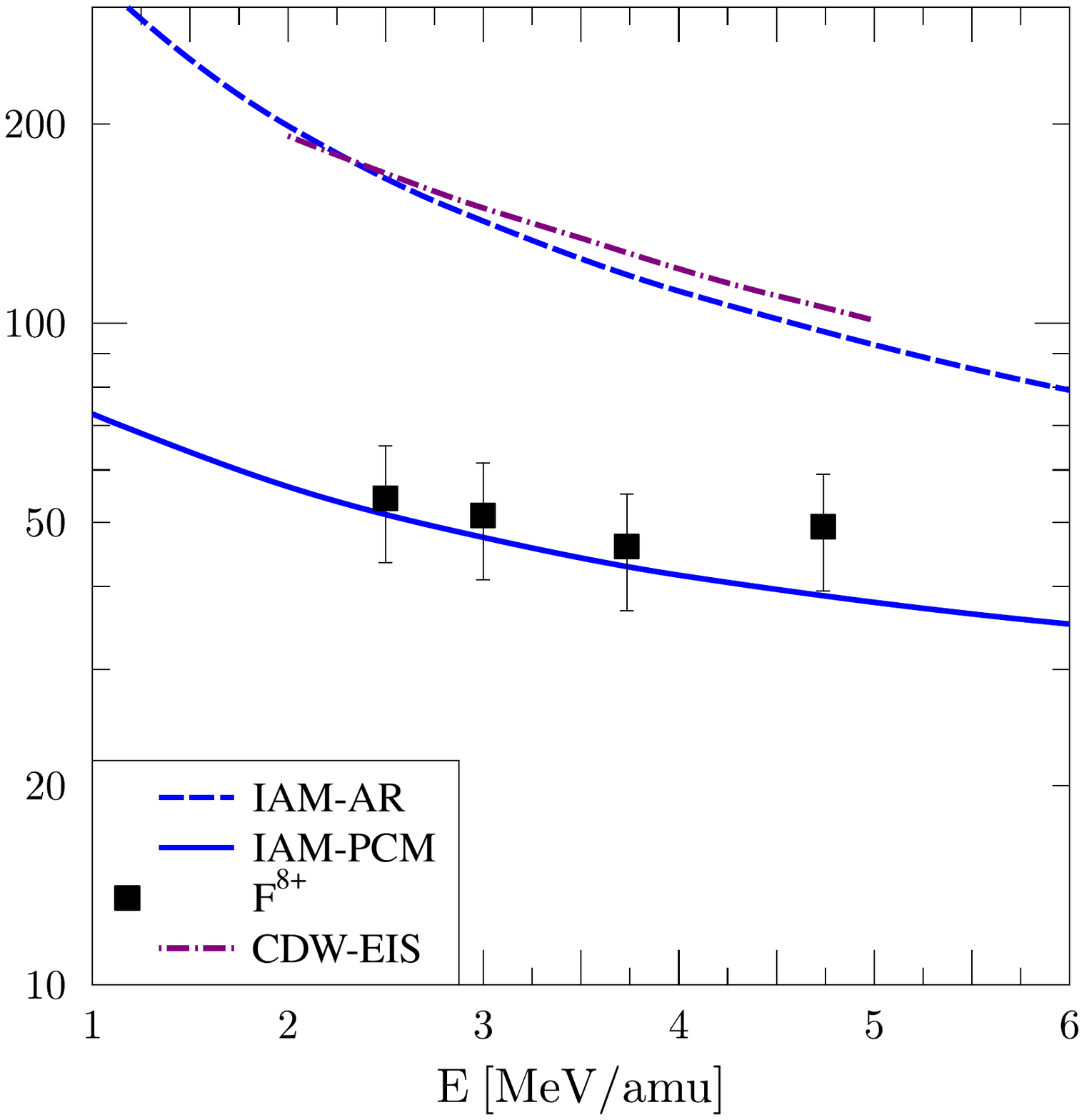}}\hspace{-1 truecm}
\end{array}$
 \caption{
   Total ionization cross sections for multiply charged ions colliding with uracil ($\rm C_4 H_4 N_2 O_2$) at high energies. The experimental data and CDW-EIS theory (dash-dotted purple lines) results
   are from Ref.~\cite{Agnihotri_2013}. The present IAM-PCM and IAM-AR results are shown as blue solid and dashed lines respectively, the 
   dotted green lines show the CTMC results of Ref.~\cite{Sarkadi_2016}).
   }
   \label{fig:Fig8}
\end{center}
\end{figure}

In order to provide more clarity concerning the comparison between theories and experiment at high energies we show in Fig.~\ref{fig:Fig8}
the data separated by projectile charge using a semi-logarithmic representation. This figure establishes how well the experimental
data are reproduced on an absolute scale by the IAM-PCM, and how far they are from the IAM-AR scaling, which matches the behavior
of the CDW-EIS and CTMC theories. The figure also demonstrates how the data really do depend on the charge state $Q$ with very little
dependence on the nuclear charge of the projectile. Some experimental data appear to deviate from the IAM-PCM results at high energies
(as commented upon before), notably for $\rm C^{5+}$, but no such deviation is visible in the same energy range for $\rm O^{7+}$. 
It remains an open question whether there are physics reasons for such a deviation from the proposed scaling behavior for particular projectiles or not.

\subsection{Collisions with water vapor ($\rm  H_2O$)}
\label{sec:expt2}

\begin{figure}
\begin{center}
%\resizebox{0.6\textwidth}{!}{%
\resizebox{0.6\textwidth}{!}{\includegraphics{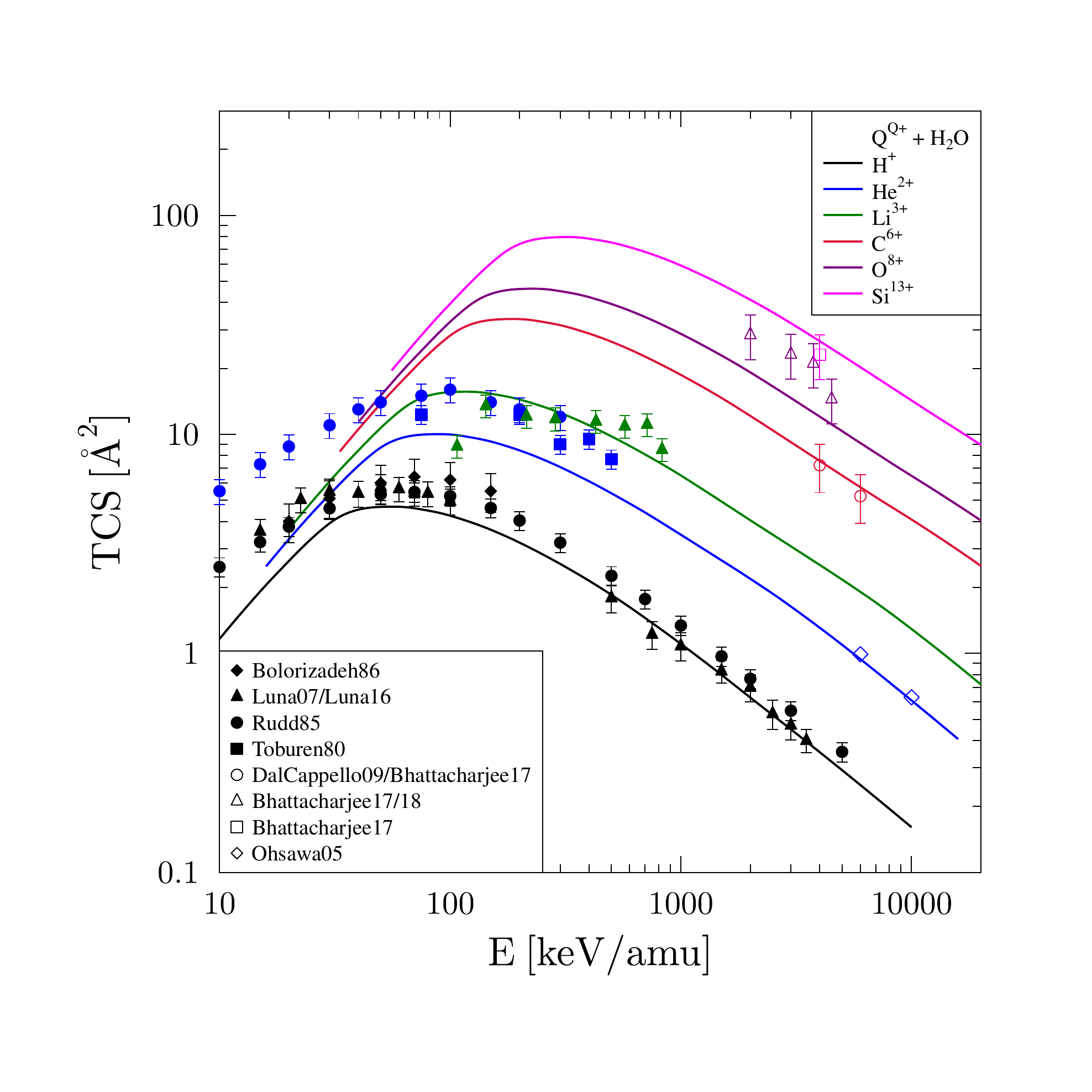}}  

\caption{
Net ionization cross sections for projectiles with charges $Q=1, 2, 3, 6, 8, 13$  impinging on water vapor. 
   Solid lines are the scaled IAM-PCM results (except for $Q=1$). 
   The experimental data are for: $Q=1$ shown as black solid squares~\cite{Toburen_80}, circles~\cite{Rudd85c}, triangles~\cite{Luna07}, diamonds~\cite{Bolorizadeh86};
   $Q=2$ shown as blue solid squares~\cite{Toburen_80}, circles~\cite{PhysRevA.32.2128}, open diamonds~\cite{Ohsawa2005}; $Q=3$ shown as green solid triangles~\cite{Luna16}; $Q=6$ shown as open red circles (6 MeV/amu):~\cite{DALCAPPELLO2009781}, (4 MeV/amu):~\cite{Bhattacharjee_2017};
   $Q=8$ shown as open purple triangles~\cite{Bhattacharjee_2016,Bhattacharjee_2018}; and $Q=13$ shown as an open magenta square~\cite{Bhattacharjee_2017}.
   }
   \label{fig:Fig9}
\end{center}
\end{figure}

In Fig.~\ref{fig:Fig9} the IAM-PCM data for proton impact, and the scaled IAM-PCM data for higher $Q$ are compared to available measurements.
The proton impact data (shown as a black curve) which form the input for the scaled cross sections are consistent in shape with the three available
sets of measurements~\cite{Toburen_80,PhysRevA.32.2128,Bolorizadeh86,Luna07} for impact energies $E>30 \ \rm keV/amu$.
They agree very well in magnitude with the data of Luna {\it et al}~\cite{Luna07}, and are a bit lower than the other data.

For smaller energies $E<30 \ \rm keV$ they do not follow the experimental data but drop faster as $E$ decreases
- a property they share with another non-perturbative, albeit classical 
calculation~\cite{PhysRevA.83.052704}. 
At these lower energies 
the IAM cross sections drop not only more rapidly than the experimental data, but also than TC-BGM ion-molecule calculations~\cite{PhysRevA.85.052704},
and even more so than the molecular-orbital based semiclassical calculation of Ref.~\cite{PhysRevA.87.032709}, which exceeds the experimental data in this range.
One may draw the conclusion that molecular effects due to the orbital energy level structure are unlikely to be captured by an IAM.

The scaled IAM-PCM net ionization cross sections for $Q=2$ (shown in blue) display a similar behavior, and fall below the experimental data
for $\rm He^{2+}$ more
markedly for $E<80 \ \rm keV/amu$. Nevertheless,
it should be noted that the position of the maximum in the cross section agrees well. 
Again, the TC-BGM ion-molecule calculation~\cite{Pausz_2014} does not suffer from this drop at low $E$. At the highest energies the IAM-PCM results
match very well with the two measured points of Ref.~\cite{Ohsawa2005}.

For $Q=3$ (shown in green) there is agreement
with the $\rm Li^{3+}$ experimental data concerning the magnitude of the cross section for for $E>80 \ \rm keV/amu$, while the shape of the
experimental data pattern differs somewhat from the scaled IAM-PCM results. The fact that the experimental data for  $\rm He^{2+}$ and $\rm Li^{3+}$ projectiles
show such similar magnitudes cannot be explained with scaling behavior, and should be resolved by further experimentation.
At the lowest energies the experiments of Luna {\it et al}~\cite{Luna16} may suffer from a shortfall due to transfer ionization 
processes yielding two protons as fragments (only one of them would be detected).

For higher-$Q$ projectiles comparison with the experimental data for $\rm C^{6+}$~\cite{Bhattacharjee_2017,Bhattacharjee_2018,DALCAPPELLO2009781},
$\rm O^{8+}$~\cite{Bhattacharjee_2016,Bhattacharjee_2018}, and $\rm Si^{13+}$~\cite{Bhattacharjee_2017} is 
very satisfactory with some tension with the $\rm O^{8+}$ data. Overall we find that the proposed IAM-PCM scaling with 
projectile charge and energy works well for the water molecule target and is strongly supported by experiment for medium to high energies.
Note that for $Q \ge 3$ the IAM-PCM (and experimental) data have not yet
reached the $Q^2$-proportional scaling of the Bethe-Born limit at the right end of Fig.~\ref{fig:Fig9}.

\begin{figure}
\begin{center}
%\resizebox{0.6\textwidth}{!}{%
\resizebox{0.6\textwidth}{!}{\includegraphics{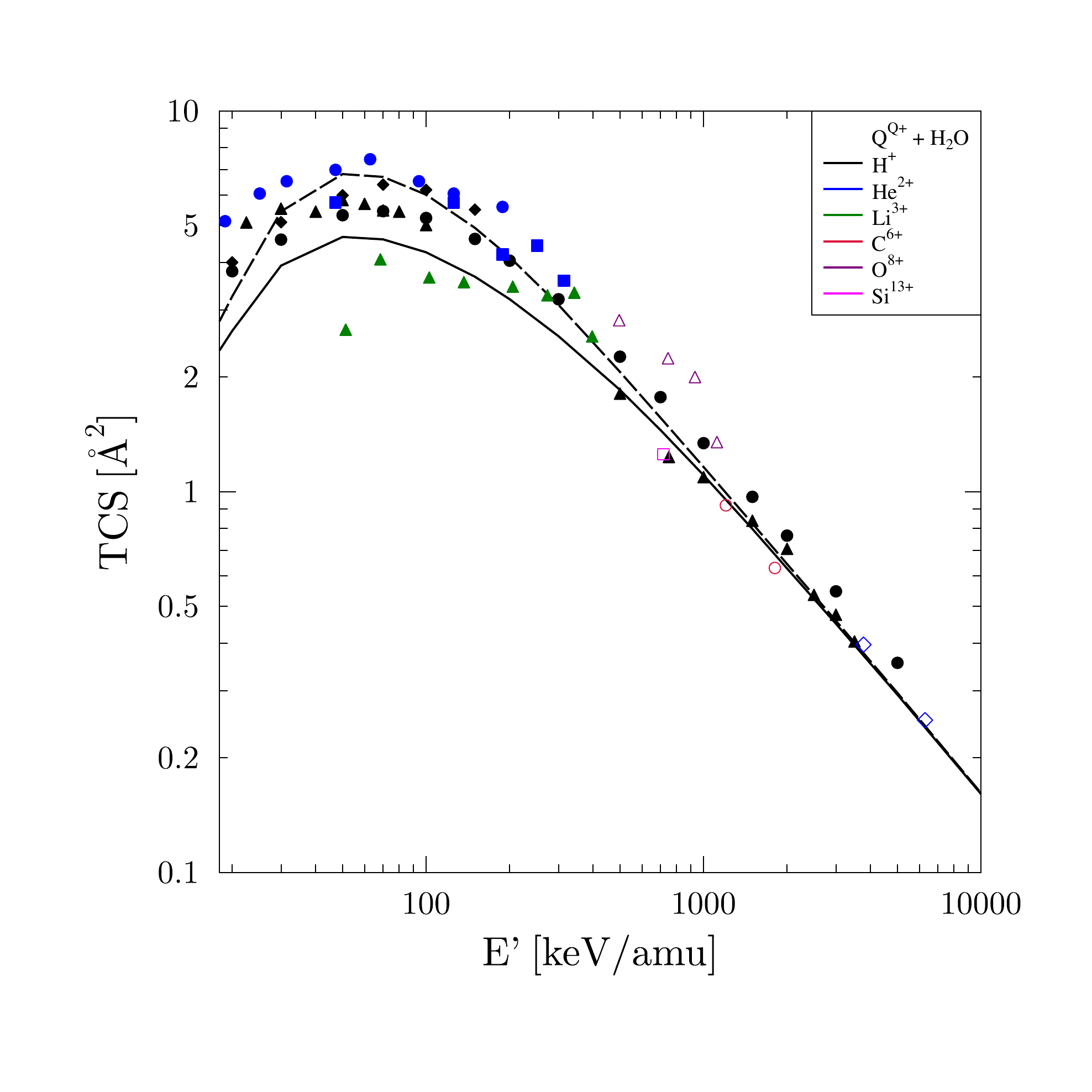}}  

\caption{
Reduced ($Q=1$) net ionization cross section obtained from experimental data for multiply charged ions colliding with $\rm H_2O$
which are obtained on the basis of the simplified IAM-PCM scaling formula (\ref{eq:eq7}). 
   The solid and dashed lines are the IAM-PCM and IAM-AR results for $Q=1$ respectively. 
   The experimental data are from the sources cited with Fig.~\ref{fig:Fig9} and share the symbol patterns, but uncertainties are omitted here.
 }
   \label{fig:Fig10}
\end{center}
\end{figure}

In Fig.~\ref{fig:Fig10} the IAM-PCM scaling behavior of Eq.~(\ref{eq:eq7}) is tested and compared against the IAM-PCM calculation for $Q=1$, while the IAM-AR
result is provided for reference. As was the case for the uracil target (Fig.~\ref{fig:UrazilReduced}) the experimental data at high energies show scatter, but follow the 
trend well - especially since there are discrepancies for different data taken for identical projectiles. At energies below $200 \ \rm keV/amu$ the two IAM results
separate as one approaches the maximum. The $\rm Li^{3+}$ impact data~\cite{Luna16} which were compared against TC-BGM ion-molecule calculations
(and which are also compared with in the recent CTMC mean-field model work of Ref.~\cite{jorge2020multicharged}) provide some support for the proposed
scaling.  At energies below $E' < 50 \ \rm keV/amu$ one may want to be critical of the IAM approach and of CTMC, since these models cannot describe the molecular effects
associated with the electronic structure of the particular molecule in question. At higher energies one can think of IAM-AR results as representing an upper bound to the reduced cross
section, while IAM-PCM may be on the low side, since the effect of the overlapping cross sectional areas does reduce the contributions from certain multiple processes, 
namely those associated with separate constituent atoms.
The reduction of multiple electron contributions may be viewed as somewhat similar to what is contained in the dynamical mean-field CTMC model~\cite{jorge2020multicharged},
which in contrast to the frozen-potential CTMC calculations shows comparable behavior to what we find in the IAM-PCM results.

\section{Conclusions}
\label{sec:conclusions}

We have presented new basis generator method calculations for scattering of projectile ions with charges $Q=2,3$ from fundamental atoms that are
the building blocks for biologically relevant molecules. The calculated net ionization cross sections were then used to compute independent atom model
cross sections for molecular targets using both the simple additivity rule (IAM-AR, which should represent the correct high-energy limit) and an independent atom model
that is based on the idea that the projectiles only experience the overlap of atomic cross sections taken as geometric areas (carried out by a pixel counting method or PCM).
The scaling behavior of the IAM-AR and IAM-PCM results is established, and a parametrization is presented that is capable of reproducing the computed IAM-PCM 
net ionization cross sections in terms of the proton-molecule IAM-PCM and IAM-AR results. To a good approximation this can also be obtained in terms of the IAM-PCM
proton-molecule cross section alone. These scaled results are then compared to experimental data for two targets, namely uracil and water vapor.
While the agreement is not perfect, a number of trends are reproduced better than by any other theory applied previously to the problem. 
The theoretical scaling model was used to compute reduced $Q=1$ cross sections from $Q \ge 2$ experimental data for uracil and water and then compared
to the proton impact case. We also pointed
out a few inconsistencies in the experimental data - this fact implies that further experiments are needed. 
For the uracil target, in particular, systematic measurements for at least $Q=1, 2$ in the tens of keV/amu  energy range would help to establish 
the entire domain where the scaling model applies in nature. Such investigations together with the presented results
will pave the way towards reliable simulations for cancer therapy with highly charged ions.

\begin{acknowledgments}
We acknowledge discussions with Alba Jorge.
We would like to thank the Center for Scientific Computing, University of Frankfurt for making their
High Performance Computing facilities available.
Financial support from the Natural Sciences and Engineering Research Council of Canada is gratefully acknowledged. \end{acknowledgments}

%
% BibTeX users please use
%\bibliographystyle{plain}
%\bibliographystyle{unsrt}

\bibliography{ChargedProjectilesRT4v2}

\begin{thebibliography}{49}%
\makeatletter
\providecommand \@ifxundefined [1]{%
 \@ifx{#1\undefined}
}%
\providecommand \@ifnum [1]{%
 \ifnum #1\expandafter \@firstoftwo
 \else \expandafter \@secondoftwo
 \fi
}%
\providecommand \@ifx [1]{%
 \ifx #1\expandafter \@firstoftwo
 \else \expandafter \@secondoftwo
 \fi
}%
\providecommand \natexlab [1]{#1}%
\providecommand \enquote  [1]{``#1''}%
\providecommand \bibnamefont  [1]{#1}%
\providecommand \bibfnamefont [1]{#1}%
\providecommand \citenamefont [1]{#1}%
\providecommand \href@noop [0]{\@secondoftwo}%
\providecommand \href [0]{\begingroup \@sanitize@url \@href}%
\providecommand \@href[1]{\@@startlink{#1}\@@href}%
\providecommand \@@href[1]{\endgroup#1\@@endlink}%
\providecommand \@sanitize@url [0]{\catcode `\\12\catcode `\$12\catcode
  `\&12\catcode `\#12\catcode `\^12\catcode `\_12\catcode `\%12\relax}%
\providecommand \@@startlink[1]{}%
\providecommand \@@endlink[0]{}%
\providecommand \url  [0]{\begingroup\@sanitize@url \@url }%
\providecommand \@url [1]{\endgroup\@href {#1}{\urlprefix }}%
\providecommand \urlprefix  [0]{URL }%
\providecommand \Eprint [0]{\href }%
\providecommand \doibase [0]{http://dx.doi.org/}%
\providecommand \selectlanguage [0]{\@gobble}%
\providecommand \bibinfo  [0]{\@secondoftwo}%
\providecommand \bibfield  [0]{\@secondoftwo}%
\providecommand \translation [1]{[#1]}%
\providecommand \BibitemOpen [0]{}%
\providecommand \bibitemStop [0]{}%
\providecommand \bibitemNoStop [0]{.\EOS\space}%
\providecommand \EOS [0]{\spacefactor3000\relax}%
\providecommand \BibitemShut  [1]{\csname bibitem#1\endcsname}%
\let\auto@bib@innerbib\@empty
%</preamble>
\bibitem [{\citenamefont {Alcocer-{\'A}vila}\ \emph {et~al.}(2019)\citenamefont
  {Alcocer-{\'A}vila}, \citenamefont {Quinto}, \citenamefont {Monti},
  \citenamefont {Rivarola},\ and\ \citenamefont {Champion}}]{Alcocer2019}%
  \BibitemOpen
  \bibfield  {author} {\bibinfo {author} {\bibfnamefont {M.~E.}\ \bibnamefont
  {Alcocer-{\'A}vila}}, \bibinfo {author} {\bibfnamefont {M.~A.}\ \bibnamefont
  {Quinto}}, \bibinfo {author} {\bibfnamefont {J.~M.}\ \bibnamefont {Monti}},
  \bibinfo {author} {\bibfnamefont {R.~D.}\ \bibnamefont {Rivarola}}, \ and\
  \bibinfo {author} {\bibfnamefont {C.}~\bibnamefont {Champion}},\ }\href
  {\doibase 10.1038/s41598-019-50270-5} {\bibfield  {journal} {\bibinfo
  {journal} {Scientific Reports}\ }\textbf {\bibinfo {volume} {9}},\ \bibinfo
  {pages} {14030} (\bibinfo {year} {2019})}\BibitemShut {NoStop}%
\bibitem [{\citenamefont {Agnihotri}\ \emph {et~al.}(2012)\citenamefont
  {Agnihotri}, \citenamefont {Kasthurirangan}, \citenamefont {Nandi},
  \citenamefont {Kumar}, \citenamefont {Galassi}, \citenamefont {Rivarola},
  \citenamefont {Foj\'on}, \citenamefont {Champion}, \citenamefont {Hanssen},
  \citenamefont {Lekadir}, \citenamefont {Weck},\ and\ \citenamefont
  {Tribedi}}]{PhysRevA.85.032711}%
  \BibitemOpen
  \bibfield  {author} {\bibinfo {author} {\bibfnamefont {A.~N.}\ \bibnamefont
  {Agnihotri}}, \bibinfo {author} {\bibfnamefont {S.}~\bibnamefont
  {Kasthurirangan}}, \bibinfo {author} {\bibfnamefont {S.}~\bibnamefont
  {Nandi}}, \bibinfo {author} {\bibfnamefont {A.}~\bibnamefont {Kumar}},
  \bibinfo {author} {\bibfnamefont {M.~E.}\ \bibnamefont {Galassi}}, \bibinfo
  {author} {\bibfnamefont {R.~D.}\ \bibnamefont {Rivarola}}, \bibinfo {author}
  {\bibfnamefont {O.}~\bibnamefont {Foj\'on}}, \bibinfo {author} {\bibfnamefont
  {C.}~\bibnamefont {Champion}}, \bibinfo {author} {\bibfnamefont
  {J.}~\bibnamefont {Hanssen}}, \bibinfo {author} {\bibfnamefont
  {H.}~\bibnamefont {Lekadir}}, \bibinfo {author} {\bibfnamefont {P.~F.}\
  \bibnamefont {Weck}}, \ and\ \bibinfo {author} {\bibfnamefont {L.~C.}\
  \bibnamefont {Tribedi}},\ }\href {\doibase 10.1103/PhysRevA.85.032711}
  {\bibfield  {journal} {\bibinfo  {journal} {Phys. Rev. A}\ }\textbf {\bibinfo
  {volume} {85}},\ \bibinfo {pages} {032711} (\bibinfo {year}
  {2012})}\BibitemShut {NoStop}%
\bibitem [{\citenamefont {Agnihotri}\ \emph {et~al.}(2013)\citenamefont
  {Agnihotri}, \citenamefont {Kasthurirangan}, \citenamefont {Nandi},
  \citenamefont {Kumar}, \citenamefont {Champion}, \citenamefont {Lekadir},
  \citenamefont {Hanssen}, \citenamefont {Weck}, \citenamefont {Galassi},
  \citenamefont {Rivarola}, \citenamefont {Foj{\'{o}}n},\ and\ \citenamefont
  {Tribedi}}]{Agnihotri_2013}%
  \BibitemOpen
  \bibfield  {author} {\bibinfo {author} {\bibfnamefont {A.~N.}\ \bibnamefont
  {Agnihotri}}, \bibinfo {author} {\bibfnamefont {S.}~\bibnamefont
  {Kasthurirangan}}, \bibinfo {author} {\bibfnamefont {S.}~\bibnamefont
  {Nandi}}, \bibinfo {author} {\bibfnamefont {A.}~\bibnamefont {Kumar}},
  \bibinfo {author} {\bibfnamefont {C.}~\bibnamefont {Champion}}, \bibinfo
  {author} {\bibfnamefont {H.}~\bibnamefont {Lekadir}}, \bibinfo {author}
  {\bibfnamefont {J.}~\bibnamefont {Hanssen}}, \bibinfo {author} {\bibfnamefont
  {P.~F.}\ \bibnamefont {Weck}}, \bibinfo {author} {\bibfnamefont {M.~E.}\
  \bibnamefont {Galassi}}, \bibinfo {author} {\bibfnamefont {R.~D.}\
  \bibnamefont {Rivarola}}, \bibinfo {author} {\bibfnamefont {O.}~\bibnamefont
  {Foj{\'{o}}n}}, \ and\ \bibinfo {author} {\bibfnamefont {L.~C.}\ \bibnamefont
  {Tribedi}},\ }\href {\doibase 10.1088/0953-4075/46/18/185201} {\bibfield
  {journal} {\bibinfo  {journal} {Journal of Physics B: Atomic, Molecular and
  Optical Physics}\ }\textbf {\bibinfo {volume} {46}},\ \bibinfo {pages}
  {185201} (\bibinfo {year} {2013})}\BibitemShut {NoStop}%
\bibitem [{\citenamefont {L\'opez-Tarifa}\ \emph {et~al.}(2011)\citenamefont
  {L\'opez-Tarifa}, \citenamefont {Herv\'e~du Penhoat}, \citenamefont
  {Vuilleumier}, \citenamefont {Gaigeot}, \citenamefont {Tavernelli},
  \citenamefont {Le~Padellec}, \citenamefont {Champeaux}, \citenamefont
  {Alcam\'{\i}}, \citenamefont {Moretto-Capelle}, \citenamefont {Mart\'{\i}n},\
  and\ \citenamefont {Politis}}]{PhysRevLett.107.023202}%
  \BibitemOpen
  \bibfield  {author} {\bibinfo {author} {\bibfnamefont {P.}~\bibnamefont
  {L\'opez-Tarifa}}, \bibinfo {author} {\bibfnamefont {M.-A.}\ \bibnamefont
  {Herv\'e~du Penhoat}}, \bibinfo {author} {\bibfnamefont {R.}~\bibnamefont
  {Vuilleumier}}, \bibinfo {author} {\bibfnamefont {M.-P.}\ \bibnamefont
  {Gaigeot}}, \bibinfo {author} {\bibfnamefont {I.}~\bibnamefont {Tavernelli}},
  \bibinfo {author} {\bibfnamefont {A.}~\bibnamefont {Le~Padellec}}, \bibinfo
  {author} {\bibfnamefont {J.-P.}\ \bibnamefont {Champeaux}}, \bibinfo {author}
  {\bibfnamefont {M.}~\bibnamefont {Alcam\'{\i}}}, \bibinfo {author}
  {\bibfnamefont {P.}~\bibnamefont {Moretto-Capelle}}, \bibinfo {author}
  {\bibfnamefont {F.}~\bibnamefont {Mart\'{\i}n}}, \ and\ \bibinfo {author}
  {\bibfnamefont {M.-F.}\ \bibnamefont {Politis}},\ }\href {\doibase
  10.1103/PhysRevLett.107.023202} {\bibfield  {journal} {\bibinfo  {journal}
  {Phys. Rev. Lett.}\ }\textbf {\bibinfo {volume} {107}},\ \bibinfo {pages}
  {023202} (\bibinfo {year} {2011})}\BibitemShut {NoStop}%
\bibitem [{\citenamefont {Covington}\ \emph {et~al.}(2017)\citenamefont
  {Covington}, \citenamefont {Hartig}, \citenamefont {Russakoff}, \citenamefont
  {Kulpins},\ and\ \citenamefont {Varga}}]{Covington17}%
  \BibitemOpen
  \bibfield  {author} {\bibinfo {author} {\bibfnamefont {C.}~\bibnamefont
  {Covington}}, \bibinfo {author} {\bibfnamefont {K.}~\bibnamefont {Hartig}},
  \bibinfo {author} {\bibfnamefont {A.}~\bibnamefont {Russakoff}}, \bibinfo
  {author} {\bibfnamefont {R.}~\bibnamefont {Kulpins}}, \ and\ \bibinfo
  {author} {\bibfnamefont {K.}~\bibnamefont {Varga}},\ }\href {\doibase
  10.1103/PhysRevA.95.052701} {\bibfield  {journal} {\bibinfo  {journal} {Phys.
  Rev. A}\ }\textbf {\bibinfo {volume} {95}},\ \bibinfo {pages} {052701}
  (\bibinfo {year} {2017})}\BibitemShut {NoStop}%
\bibitem [{\citenamefont {Salo}\ \emph {et~al.}(2018)\citenamefont {Salo},
  \citenamefont {Alberg-Fl\o{}jborg},\ and\ \citenamefont
  {Solov'yov}}]{Salo18}%
  \BibitemOpen
  \bibfield  {author} {\bibinfo {author} {\bibfnamefont {A.~B.}\ \bibnamefont
  {Salo}}, \bibinfo {author} {\bibfnamefont {A.}~\bibnamefont
  {Alberg-Fl\o{}jborg}}, \ and\ \bibinfo {author} {\bibfnamefont {I.~A.}\
  \bibnamefont {Solov'yov}},\ }\href {\doibase 10.1103/PhysRevA.98.012702}
  {\bibfield  {journal} {\bibinfo  {journal} {Phys. Rev. A}\ }\textbf {\bibinfo
  {volume} {98}},\ \bibinfo {pages} {012702} (\bibinfo {year}
  {2018})}\BibitemShut {NoStop}%
\bibitem [{\citenamefont {Itoh}\ \emph {et~al.}(2013)\citenamefont {Itoh},
  \citenamefont {Iriki}, \citenamefont {Imai}, \citenamefont {Champion},\ and\
  \citenamefont {Rivarola}}]{Itoh13}%
  \BibitemOpen
  \bibfield  {author} {\bibinfo {author} {\bibfnamefont {A.}~\bibnamefont
  {Itoh}}, \bibinfo {author} {\bibfnamefont {Y.}~\bibnamefont {Iriki}},
  \bibinfo {author} {\bibfnamefont {M.}~\bibnamefont {Imai}}, \bibinfo {author}
  {\bibfnamefont {C.}~\bibnamefont {Champion}}, \ and\ \bibinfo {author}
  {\bibfnamefont {R.~D.}\ \bibnamefont {Rivarola}},\ }\href {\doibase
  10.1103/PhysRevA.88.052711} {\bibfield  {journal} {\bibinfo  {journal} {Phys.
  Rev. A}\ }\textbf {\bibinfo {volume} {88}},\ \bibinfo {pages} {052711}
  (\bibinfo {year} {2013})}\BibitemShut {NoStop}%
\bibitem [{\citenamefont {Paredes}\ \emph {et~al.}(2015)\citenamefont
  {Paredes}, \citenamefont {Illescas},\ and\ \citenamefont
  {M\'endez}}]{Paredes15}%
  \BibitemOpen
  \bibfield  {author} {\bibinfo {author} {\bibfnamefont {S.}~\bibnamefont
  {Paredes}}, \bibinfo {author} {\bibfnamefont {C.}~\bibnamefont {Illescas}}, \
  and\ \bibinfo {author} {\bibfnamefont {L.}~\bibnamefont {M\'endez}},\ }\href
  {\doibase 10.1140/epjd/e2015-60106-8} {\bibfield  {journal} {\bibinfo
  {journal} {Eur. Phys. J. D}\ }\textbf {\bibinfo {volume} {69}},\ \bibinfo
  {pages} {178} (\bibinfo {year} {2015})}\BibitemShut {NoStop}%
\bibitem [{\citenamefont {L\"udde}\ \emph {et~al.}(2016)\citenamefont
  {L\"udde}, \citenamefont {Achenbach}, \citenamefont {Kalkbrenner},
  \citenamefont {Jankowiak},\ and\ \citenamefont {Kirchner}}]{hjl16}%
  \BibitemOpen
  \bibfield  {author} {\bibinfo {author} {\bibfnamefont {H.~J.}\ \bibnamefont
  {L\"udde}}, \bibinfo {author} {\bibfnamefont {A.}~\bibnamefont {Achenbach}},
  \bibinfo {author} {\bibfnamefont {T.}~\bibnamefont {Kalkbrenner}}, \bibinfo
  {author} {\bibfnamefont {H.-C.}\ \bibnamefont {Jankowiak}}, \ and\ \bibinfo
  {author} {\bibfnamefont {T.}~\bibnamefont {Kirchner}},\ }\href
  {https://doi.org/10.1140/epjd/e2016-70097-5} {\bibfield  {journal} {\bibinfo
  {journal} {Eur. Phys. J. D}\ }\textbf {\bibinfo {volume} {70}},\ \bibinfo
  {pages} {82} (\bibinfo {year} {2016})}\BibitemShut {NoStop}%
\bibitem [{\citenamefont {L\"udde}\ \emph {et~al.}(2018)\citenamefont
  {L\"udde}, \citenamefont {Horbatsch},\ and\ \citenamefont
  {Kirchner}}]{hjl18}%
  \BibitemOpen
  \bibfield  {author} {\bibinfo {author} {\bibfnamefont {H.~J.}\ \bibnamefont
  {L\"udde}}, \bibinfo {author} {\bibfnamefont {M.}~\bibnamefont {Horbatsch}},
  \ and\ \bibinfo {author} {\bibfnamefont {T.}~\bibnamefont {Kirchner}},\
  }\href {https://doi.org/10.1140/epjb/e2018-90165-x} {\bibfield  {journal}
  {\bibinfo  {journal} {Eur. Phys. J. B}\ }\textbf {\bibinfo {volume} {91}},\
  \bibinfo {pages} {99} (\bibinfo {year} {2018})}\BibitemShut {NoStop}%
\bibitem [{\citenamefont {L\"udde}\ \emph {et~al.}(2019)\citenamefont
  {L\"udde}, \citenamefont {Horbatsch},\ and\ \citenamefont
  {Kirchner}}]{hjl19}%
  \BibitemOpen
  \bibfield  {author} {\bibinfo {author} {\bibfnamefont {H.~J.}\ \bibnamefont
  {L\"udde}}, \bibinfo {author} {\bibfnamefont {M.}~\bibnamefont {Horbatsch}},
  \ and\ \bibinfo {author} {\bibfnamefont {T.}~\bibnamefont {Kirchner}},\
  }\href {https://doi.org/10.1088/1361-6455/ab3a63} {\bibfield  {journal}
  {\bibinfo  {journal} {J. Phys. B}\ }\textbf {\bibinfo {volume} {52}},\
  \bibinfo {pages} {195203} (\bibinfo {year} {2019})}\BibitemShut {NoStop}%
\bibitem [{\citenamefont {L{\"u}dde}\ \emph {et~al.}(2019)\citenamefont
  {L{\"u}dde}, \citenamefont {Horbatsch},\ and\ \citenamefont
  {Kirchner}}]{hjl19b}%
  \BibitemOpen
  \bibfield  {author} {\bibinfo {author} {\bibfnamefont {H.~J.}\ \bibnamefont
  {L{\"u}dde}}, \bibinfo {author} {\bibfnamefont {M.}~\bibnamefont
  {Horbatsch}}, \ and\ \bibinfo {author} {\bibfnamefont {T.}~\bibnamefont
  {Kirchner}},\ }\href {\doibase 10.1140/epjd/e2019-100344-2} {\bibfield
  {journal} {\bibinfo  {journal} {The European Physical Journal D}\ }\textbf
  {\bibinfo {volume} {73}},\ \bibinfo {pages} {249} (\bibinfo {year}
  {2019})}\BibitemShut {NoStop}%
\bibitem [{\citenamefont {Zapukhlyak}\ \emph {et~al.}(2005)\citenamefont
  {Zapukhlyak}, \citenamefont {Kirchner}, \citenamefont {L\"udde},
  \citenamefont {Knoop}, \citenamefont {Morgenstern},\ and\ \citenamefont
  {Hoekstra}}]{tcbgm}%
  \BibitemOpen
  \bibfield  {author} {\bibinfo {author} {\bibfnamefont {M.}~\bibnamefont
  {Zapukhlyak}}, \bibinfo {author} {\bibfnamefont {T.}~\bibnamefont
  {Kirchner}}, \bibinfo {author} {\bibfnamefont {H.~J.}\ \bibnamefont
  {L\"udde}}, \bibinfo {author} {\bibfnamefont {S.}~\bibnamefont {Knoop}},
  \bibinfo {author} {\bibfnamefont {R.}~\bibnamefont {Morgenstern}}, \ and\
  \bibinfo {author} {\bibfnamefont {R.}~\bibnamefont {Hoekstra}},\ }\href
  {\doibase 10.1088/0953-4075/38/14/003} {\bibfield  {journal} {\bibinfo
  {journal} {J. Phys. B}\ }\textbf {\bibinfo {volume} {38}},\ \bibinfo {pages}
  {2353} (\bibinfo {year} {2005})}\BibitemShut {NoStop}%
\bibitem [{\citenamefont {Mendez}\ \emph {et~al.}(2019)\citenamefont {Mendez},
  \citenamefont {Montanari},\ and\ \citenamefont
  {Miraglia}}]{mendez2019ionization}%
  \BibitemOpen
  \bibfield  {author} {\bibinfo {author} {\bibfnamefont {A.~M.~P.}\
  \bibnamefont {Mendez}}, \bibinfo {author} {\bibfnamefont {C.~C.}\
  \bibnamefont {Montanari}}, \ and\ \bibinfo {author} {\bibfnamefont {J.~E.}\
  \bibnamefont {Miraglia}},\ }\href {\doibase 10.1088/1361-6455/ab6052}
  {\bibfield  {journal} {\bibinfo  {journal} {Journal of Physics B: Atomic,
  Molecular and Optical Physics}\ } (\bibinfo {year} {2019}),\
  10.1088/1361-6455/ab6052}\BibitemShut {NoStop}%
\bibitem [{\citenamefont {Mendez}\ \emph {et~al.}(2020)\citenamefont {Mendez},
  \citenamefont {Montanari},\ and\ \citenamefont
  {Miraglia}}]{alej2020universal}%
  \BibitemOpen
  \bibfield  {author} {\bibinfo {author} {\bibfnamefont {A.~M.~P.}\
  \bibnamefont {Mendez}}, \bibinfo {author} {\bibfnamefont {C.~C.}\
  \bibnamefont {Montanari}}, \ and\ \bibinfo {author} {\bibfnamefont {J.~E.}\
  \bibnamefont {Miraglia}},\ }\href@noop {} {\enquote {\bibinfo {title}
  {Universal scaling for the ionization of biological molecules by highly
  charged ions},}\ } (\bibinfo {year} {2020}),\ \Eprint
  {http://arxiv.org/abs/2003.04338} {arXiv:2003.04338 [physics.atm-clus]}
  \BibitemShut {NoStop}%
\bibitem [{\citenamefont {Jorge}\ \emph {et~al.}(2019)\citenamefont {Jorge},
  \citenamefont {Horbatsch}, \citenamefont {Illescas},\ and\ \citenamefont
  {Kirchner}}]{PhysRevA.99.062701}%
  \BibitemOpen
  \bibfield  {author} {\bibinfo {author} {\bibfnamefont {A.}~\bibnamefont
  {Jorge}}, \bibinfo {author} {\bibfnamefont {M.}~\bibnamefont {Horbatsch}},
  \bibinfo {author} {\bibfnamefont {C.}~\bibnamefont {Illescas}}, \ and\
  \bibinfo {author} {\bibfnamefont {T.}~\bibnamefont {Kirchner}},\ }\href
  {\doibase 10.1103/PhysRevA.99.062701} {\bibfield  {journal} {\bibinfo
  {journal} {Phys. Rev. A}\ }\textbf {\bibinfo {volume} {99}},\ \bibinfo
  {pages} {062701} (\bibinfo {year} {2019})}\BibitemShut {NoStop}%
\bibitem [{\citenamefont {Toburen}\ \emph {et~al.}(1980)\citenamefont
  {Toburen}, \citenamefont {Wilson},\ and\ \citenamefont
  {Popowich}}]{Toburen_80}%
  \BibitemOpen
  \bibfield  {author} {\bibinfo {author} {\bibfnamefont {L.~H.}\ \bibnamefont
  {Toburen}}, \bibinfo {author} {\bibfnamefont {W.~E.}\ \bibnamefont {Wilson}},
  \ and\ \bibinfo {author} {\bibfnamefont {R.~J.}\ \bibnamefont {Popowich}},\
  }\href {http://www.jstor.org/stable/3575234} {\bibfield  {journal} {\bibinfo
  {journal} {Radiation Research}\ }\textbf {\bibinfo {volume} {82}},\ \bibinfo
  {pages} {27} (\bibinfo {year} {1980})}\BibitemShut {NoStop}%
\bibitem [{\citenamefont {Rudd}\ \emph
  {et~al.}(1985{\natexlab{a}})\citenamefont {Rudd}, \citenamefont {Goffe},\
  and\ \citenamefont {Itoh}}]{PhysRevA.32.2128}%
  \BibitemOpen
  \bibfield  {author} {\bibinfo {author} {\bibfnamefont {M.~E.}\ \bibnamefont
  {Rudd}}, \bibinfo {author} {\bibfnamefont {T.~V.}\ \bibnamefont {Goffe}}, \
  and\ \bibinfo {author} {\bibfnamefont {A.}~\bibnamefont {Itoh}},\ }\href
  {\doibase 10.1103/PhysRevA.32.2128} {\bibfield  {journal} {\bibinfo
  {journal} {Phys. Rev. A}\ }\textbf {\bibinfo {volume} {32}},\ \bibinfo
  {pages} {2128} (\bibinfo {year} {1985}{\natexlab{a}})}\BibitemShut {NoStop}%
\bibitem [{\citenamefont {Bolorizadeh}\ and\ \citenamefont
  {Rudd}(1986)}]{Bolorizadeh86}%
  \BibitemOpen
  \bibfield  {author} {\bibinfo {author} {\bibfnamefont {M.~A.}\ \bibnamefont
  {Bolorizadeh}}\ and\ \bibinfo {author} {\bibfnamefont {M.~E.}\ \bibnamefont
  {Rudd}},\ }\href {https://link.aps.org/doi/10.1103/PhysRevA.33.888}
  {\bibfield  {journal} {\bibinfo  {journal} {Phys. Rev. A}\ }\textbf {\bibinfo
  {volume} {33}},\ \bibinfo {pages} {888} (\bibinfo {year} {1986})}\BibitemShut
  {NoStop}%
\bibitem [{\citenamefont {Luna}\ \emph {et~al.}(2007)\citenamefont {Luna},
  \citenamefont {de~Barros}, \citenamefont {Wyer}, \citenamefont {Scully},
  \citenamefont {Lecointre}, \citenamefont {Garcia}, \citenamefont {Sigaud},
  \citenamefont {Santos}, \citenamefont {Senthil}, \citenamefont {Shah},
  \citenamefont {Latimer},\ and\ \citenamefont {Montenegro}}]{Luna07}%
  \BibitemOpen
  \bibfield  {author} {\bibinfo {author} {\bibfnamefont {H.}~\bibnamefont
  {Luna}}, \bibinfo {author} {\bibfnamefont {A.~L.~F.}\ \bibnamefont
  {de~Barros}}, \bibinfo {author} {\bibfnamefont {J.~A.}\ \bibnamefont {Wyer}},
  \bibinfo {author} {\bibfnamefont {S.~W.~J.}\ \bibnamefont {Scully}}, \bibinfo
  {author} {\bibfnamefont {J.}~\bibnamefont {Lecointre}}, \bibinfo {author}
  {\bibfnamefont {P.~M.~Y.}\ \bibnamefont {Garcia}}, \bibinfo {author}
  {\bibfnamefont {G.~M.}\ \bibnamefont {Sigaud}}, \bibinfo {author}
  {\bibfnamefont {A.~C.~F.}\ \bibnamefont {Santos}}, \bibinfo {author}
  {\bibfnamefont {V.}~\bibnamefont {Senthil}}, \bibinfo {author} {\bibfnamefont
  {M.~B.}\ \bibnamefont {Shah}}, \bibinfo {author} {\bibfnamefont {C.~J.}\
  \bibnamefont {Latimer}}, \ and\ \bibinfo {author} {\bibfnamefont {E.~C.}\
  \bibnamefont {Montenegro}},\ }\href {\doibase 10.1103/PhysRevA.75.042711}
  {\bibfield  {journal} {\bibinfo  {journal} {Phys. Rev. A}\ }\textbf {\bibinfo
  {volume} {75}},\ \bibinfo {pages} {042711} (\bibinfo {year}
  {2007})}\BibitemShut {NoStop}%
\bibitem [{\citenamefont {Luna}\ \emph {et~al.}(2016)\citenamefont {Luna},
  \citenamefont {Wolff}, \citenamefont {Montenegro}, \citenamefont {Tavares},
  \citenamefont {L\"udde}, \citenamefont {Schenk}, \citenamefont {Horbatsch},\
  and\ \citenamefont {Kirchner}}]{Luna16}%
  \BibitemOpen
  \bibfield  {author} {\bibinfo {author} {\bibfnamefont {H.}~\bibnamefont
  {Luna}}, \bibinfo {author} {\bibfnamefont {W.}~\bibnamefont {Wolff}},
  \bibinfo {author} {\bibfnamefont {E.~C.}\ \bibnamefont {Montenegro}},
  \bibinfo {author} {\bibfnamefont {A.~C.}\ \bibnamefont {Tavares}}, \bibinfo
  {author} {\bibfnamefont {H.~J.}\ \bibnamefont {L\"udde}}, \bibinfo {author}
  {\bibfnamefont {G.}~\bibnamefont {Schenk}}, \bibinfo {author} {\bibfnamefont
  {M.}~\bibnamefont {Horbatsch}}, \ and\ \bibinfo {author} {\bibfnamefont
  {T.}~\bibnamefont {Kirchner}},\ }\href {\doibase 10.1103/PhysRevA.93.052705}
  {\bibfield  {journal} {\bibinfo  {journal} {Phys. Rev. A}\ }\textbf {\bibinfo
  {volume} {93}},\ \bibinfo {pages} {052705} (\bibinfo {year}
  {2016})}\BibitemShut {NoStop}%
\bibitem [{\citenamefont {Ohsawa}\ \emph {et~al.}(2013)\citenamefont {Ohsawa},
  \citenamefont {Tawara}, \citenamefont {Soga}, \citenamefont {Galassi},\ and\
  \citenamefont {Rivarola}}]{Ohsawa_2013}%
  \BibitemOpen
  \bibfield  {author} {\bibinfo {author} {\bibfnamefont {D.}~\bibnamefont
  {Ohsawa}}, \bibinfo {author} {\bibfnamefont {H.}~\bibnamefont {Tawara}},
  \bibinfo {author} {\bibfnamefont {F.}~\bibnamefont {Soga}}, \bibinfo {author}
  {\bibfnamefont {M.~E.}\ \bibnamefont {Galassi}}, \ and\ \bibinfo {author}
  {\bibfnamefont {R.~D.}\ \bibnamefont {Rivarola}},\ }\href
  {https://doi.org/10.1088%2F0031-8949%2F2013%2Ft156%2F014039} {\bibfield
  {journal} {\bibinfo  {journal} {Physica Scripta}\ }\textbf {\bibinfo {volume}
  {T156}},\ \bibinfo {pages} {014039} (\bibinfo {year} {2013})}\BibitemShut
  {NoStop}%
\bibitem [{\citenamefont {Bhattacharjee}\ \emph {et~al.}(2016)\citenamefont
  {Bhattacharjee}, \citenamefont {Biswas}, \citenamefont {Bagdia},
  \citenamefont {Roychowdhury}, \citenamefont {Nandi}, \citenamefont {Misra},
  \citenamefont {Monti}, \citenamefont {Tachino}, \citenamefont {Rivarola},
  \citenamefont {Champion},\ and\ \citenamefont
  {Tribedi}}]{Bhattacharjee_2016}%
  \BibitemOpen
  \bibfield  {author} {\bibinfo {author} {\bibfnamefont {S.}~\bibnamefont
  {Bhattacharjee}}, \bibinfo {author} {\bibfnamefont {S.}~\bibnamefont
  {Biswas}}, \bibinfo {author} {\bibfnamefont {C.}~\bibnamefont {Bagdia}},
  \bibinfo {author} {\bibfnamefont {M.}~\bibnamefont {Roychowdhury}}, \bibinfo
  {author} {\bibfnamefont {S.}~\bibnamefont {Nandi}}, \bibinfo {author}
  {\bibfnamefont {D.}~\bibnamefont {Misra}}, \bibinfo {author} {\bibfnamefont
  {J.~M.}\ \bibnamefont {Monti}}, \bibinfo {author} {\bibfnamefont {C.~A.}\
  \bibnamefont {Tachino}}, \bibinfo {author} {\bibfnamefont {R.~D.}\
  \bibnamefont {Rivarola}}, \bibinfo {author} {\bibfnamefont {C.}~\bibnamefont
  {Champion}}, \ and\ \bibinfo {author} {\bibfnamefont {L.~C.}\ \bibnamefont
  {Tribedi}},\ }\href {https://doi.org/10.1088%2F0953-4075%2F49%2F6%2F065202}
  {\bibfield  {journal} {\bibinfo  {journal} {Journal of Physics B: Atomic,
  Molecular and Optical Physics}\ }\textbf {\bibinfo {volume} {49}},\ \bibinfo
  {pages} {065202} (\bibinfo {year} {2016})}\BibitemShut {NoStop}%
\bibitem [{\citenamefont {Bhattacharjee}\ \emph {et~al.}(2017)\citenamefont
  {Bhattacharjee}, \citenamefont {Biswas}, \citenamefont {Monti}, \citenamefont
  {Rivarola},\ and\ \citenamefont {Tribedi}}]{Bhattacharjee_2017}%
  \BibitemOpen
  \bibfield  {author} {\bibinfo {author} {\bibfnamefont {S.}~\bibnamefont
  {Bhattacharjee}}, \bibinfo {author} {\bibfnamefont {S.}~\bibnamefont
  {Biswas}}, \bibinfo {author} {\bibfnamefont {J.~M.}\ \bibnamefont {Monti}},
  \bibinfo {author} {\bibfnamefont {R.~D.}\ \bibnamefont {Rivarola}}, \ and\
  \bibinfo {author} {\bibfnamefont {L.~C.}\ \bibnamefont {Tribedi}},\ }\href
  {\doibase 10.1103/PhysRevA.96.052707} {\bibfield  {journal} {\bibinfo
  {journal} {Phys. Rev. A}\ }\textbf {\bibinfo {volume} {96}},\ \bibinfo
  {pages} {052707} (\bibinfo {year} {2017})}\BibitemShut {NoStop}%
\bibitem [{\citenamefont {{Bhattacharjee, Shamik}}\ \emph
  {et~al.}(2018)\citenamefont {{Bhattacharjee, Shamik}}, \citenamefont
  {{Bagdia, Chandan}}, \citenamefont {{Chowdhury, Madhusree Roy}},
  \citenamefont {{Monti, Juan M.}}, \citenamefont {{Rivarola, Roberto D.}},\
  and\ \citenamefont {{Tribedi, Lokesh C.}}}]{Bhattacharjee_2018}%
  \BibitemOpen
  \bibfield  {author} {\bibinfo {author} {\bibnamefont {{Bhattacharjee,
  Shamik}}}, \bibinfo {author} {\bibnamefont {{Bagdia, Chandan}}}, \bibinfo
  {author} {\bibnamefont {{Chowdhury, Madhusree Roy}}}, \bibinfo {author}
  {\bibnamefont {{Monti, Juan M.}}}, \bibinfo {author} {\bibnamefont
  {{Rivarola, Roberto D.}}}, \ and\ \bibinfo {author} {\bibnamefont {{Tribedi,
  Lokesh C.}}},\ }\href {\doibase 10.1140/epjd/e2017-80265-8} {\bibfield
  {journal} {\bibinfo  {journal} {Eur. Phys. J. D}\ }\textbf {\bibinfo {volume}
  {72}},\ \bibinfo {pages} {15} (\bibinfo {year} {2018})}\BibitemShut {NoStop}%
\bibitem [{\citenamefont {L\"udde}\ \emph {et~al.}(2009)\citenamefont
  {L\"udde}, \citenamefont {Spranger}, \citenamefont {Horbatsch},\ and\
  \citenamefont {Kirchner}}]{hjl09}%
  \BibitemOpen
  \bibfield  {author} {\bibinfo {author} {\bibfnamefont {H.~J.}\ \bibnamefont
  {L\"udde}}, \bibinfo {author} {\bibfnamefont {T.}~\bibnamefont {Spranger}},
  \bibinfo {author} {\bibfnamefont {M.}~\bibnamefont {Horbatsch}}, \ and\
  \bibinfo {author} {\bibfnamefont {T.}~\bibnamefont {Kirchner}},\ }\href
  {https://link.aps.org/doi/10.1103/PhysRevA.80.060702} {\bibfield  {journal}
  {\bibinfo  {journal} {Phys. Rev. A}\ }\textbf {\bibinfo {volume} {80}},\
  \bibinfo {pages} {060702(R)} (\bibinfo {year} {2009})}\BibitemShut {NoStop}%
\bibitem [{\citenamefont {Murakami}\ \emph
  {et~al.}(2012{\natexlab{a}})\citenamefont {Murakami}, \citenamefont
  {Kirchner}, \citenamefont {Horbatsch},\ and\ \citenamefont
  {L\"udde}}]{PhysRevA.85.052704}%
  \BibitemOpen
  \bibfield  {author} {\bibinfo {author} {\bibfnamefont {M.}~\bibnamefont
  {Murakami}}, \bibinfo {author} {\bibfnamefont {T.}~\bibnamefont {Kirchner}},
  \bibinfo {author} {\bibfnamefont {M.}~\bibnamefont {Horbatsch}}, \ and\
  \bibinfo {author} {\bibfnamefont {H.~J.}\ \bibnamefont {L\"udde}},\ }\href
  {\doibase 10.1103/PhysRevA.85.052704} {\bibfield  {journal} {\bibinfo
  {journal} {Phys. Rev. A}\ }\textbf {\bibinfo {volume} {85}},\ \bibinfo
  {pages} {052704} (\bibinfo {year} {2012}{\natexlab{a}})}\BibitemShut
  {NoStop}%
\bibitem [{\citenamefont {Murakami}\ \emph
  {et~al.}(2012{\natexlab{b}})\citenamefont {Murakami}, \citenamefont
  {Kirchner}, \citenamefont {Horbatsch},\ and\ \citenamefont
  {L\"udde}}]{PhysRevA.85.052713}%
  \BibitemOpen
  \bibfield  {author} {\bibinfo {author} {\bibfnamefont {M.}~\bibnamefont
  {Murakami}}, \bibinfo {author} {\bibfnamefont {T.}~\bibnamefont {Kirchner}},
  \bibinfo {author} {\bibfnamefont {M.}~\bibnamefont {Horbatsch}}, \ and\
  \bibinfo {author} {\bibfnamefont {H.~J.}\ \bibnamefont {L\"udde}},\ }\href
  {\doibase 10.1103/PhysRevA.85.052713} {\bibfield  {journal} {\bibinfo
  {journal} {Phys. Rev. A}\ }\textbf {\bibinfo {volume} {85}},\ \bibinfo
  {pages} {052713} (\bibinfo {year} {2012}{\natexlab{b}})}\BibitemShut
  {NoStop}%
\bibitem [{\citenamefont {Murakami}\ \emph
  {et~al.}(2012{\natexlab{c}})\citenamefont {Murakami}, \citenamefont
  {Kirchner}, \citenamefont {Horbatsch},\ and\ \citenamefont
  {L\"udde}}]{PhysRevA.86.022719}%
  \BibitemOpen
  \bibfield  {author} {\bibinfo {author} {\bibfnamefont {M.}~\bibnamefont
  {Murakami}}, \bibinfo {author} {\bibfnamefont {T.}~\bibnamefont {Kirchner}},
  \bibinfo {author} {\bibfnamefont {M.}~\bibnamefont {Horbatsch}}, \ and\
  \bibinfo {author} {\bibfnamefont {H.~J.}\ \bibnamefont {L\"udde}},\ }\href
  {\doibase 10.1103/PhysRevA.86.022719} {\bibfield  {journal} {\bibinfo
  {journal} {Phys. Rev. A}\ }\textbf {\bibinfo {volume} {86}},\ \bibinfo
  {pages} {022719} (\bibinfo {year} {2012}{\natexlab{c}})}\BibitemShut
  {NoStop}%
\bibitem [{\citenamefont {Pausz}\ \emph {et~al.}(2014)\citenamefont {Pausz},
  \citenamefont {Lüdde}, \citenamefont {Murakami}, \citenamefont {Horbatsch},\
  and\ \citenamefont {Kirchner}}]{Pausz_2014}%
  \BibitemOpen
  \bibfield  {author} {\bibinfo {author} {\bibfnamefont {T.}~\bibnamefont
  {Pausz}}, \bibinfo {author} {\bibfnamefont {H.~J.}\ \bibnamefont {Lüdde}},
  \bibinfo {author} {\bibfnamefont {M.}~\bibnamefont {Murakami}}, \bibinfo
  {author} {\bibfnamefont {M.}~\bibnamefont {Horbatsch}}, \ and\ \bibinfo
  {author} {\bibfnamefont {T.}~\bibnamefont {Kirchner}},\ }\href {\doibase
  10.1088/1742-6596/488/10/102013} {\bibfield  {journal} {\bibinfo  {journal}
  {Journal of Physics: Conference Series}\ }\textbf {\bibinfo {volume} {488}},\
  \bibinfo {pages} {102013} (\bibinfo {year} {2014})}\BibitemShut {NoStop}%
\bibitem [{\citenamefont {Rubio}\ \emph {et~al.}(2008)\citenamefont {Rubio},
  \citenamefont {Serrano-Andrés},\ and\ \citenamefont
  {Merchán}}]{doi:10.1063/1.2837827}%
  \BibitemOpen
  \bibfield  {author} {\bibinfo {author} {\bibfnamefont {M.}~\bibnamefont
  {Rubio}}, \bibinfo {author} {\bibfnamefont {L.}~\bibnamefont
  {Serrano-Andrés}}, \ and\ \bibinfo {author} {\bibfnamefont {M.}~\bibnamefont
  {Merchán}},\ }\href {\doibase 10.1063/1.2837827} {\bibfield  {journal}
  {\bibinfo  {journal} {The Journal of Chemical Physics}\ }\textbf {\bibinfo
  {volume} {128}},\ \bibinfo {pages} {104305} (\bibinfo {year} {2008})},\
  \Eprint {http://arxiv.org/abs/https://doi.org/10.1063/1.2837827}
  {https://doi.org/10.1063/1.2837827} \BibitemShut {NoStop}%
\bibitem [{\citenamefont {Errea}\ \emph {et~al.}(2013)\citenamefont {Errea},
  \citenamefont {Illescas}, \citenamefont {M\'endez},\ and\ \citenamefont
  {Rabad\'an}}]{PhysRevA.87.032709}%
  \BibitemOpen
  \bibfield  {author} {\bibinfo {author} {\bibfnamefont {L.~F.}\ \bibnamefont
  {Errea}}, \bibinfo {author} {\bibfnamefont {C.}~\bibnamefont {Illescas}},
  \bibinfo {author} {\bibfnamefont {L.}~\bibnamefont {M\'endez}}, \ and\
  \bibinfo {author} {\bibfnamefont {I.}~\bibnamefont {Rabad\'an}},\ }\href
  {\doibase 10.1103/PhysRevA.87.032709} {\bibfield  {journal} {\bibinfo
  {journal} {Phys. Rev. A}\ }\textbf {\bibinfo {volume} {87}},\ \bibinfo
  {pages} {032709} (\bibinfo {year} {2013})}\BibitemShut {NoStop}%
\bibitem [{\citenamefont {Illescas}\ \emph {et~al.}(2011)\citenamefont
  {Illescas}, \citenamefont {Errea}, \citenamefont {M\'endez}, \citenamefont
  {Pons}, \citenamefont {Rabad\'an},\ and\ \citenamefont
  {Riera}}]{PhysRevA.83.052704}%
  \BibitemOpen
  \bibfield  {author} {\bibinfo {author} {\bibfnamefont {C.}~\bibnamefont
  {Illescas}}, \bibinfo {author} {\bibfnamefont {L.~F.}\ \bibnamefont {Errea}},
  \bibinfo {author} {\bibfnamefont {L.}~\bibnamefont {M\'endez}}, \bibinfo
  {author} {\bibfnamefont {B.}~\bibnamefont {Pons}}, \bibinfo {author}
  {\bibfnamefont {I.}~\bibnamefont {Rabad\'an}}, \ and\ \bibinfo {author}
  {\bibfnamefont {A.}~\bibnamefont {Riera}},\ }\href {\doibase
  10.1103/PhysRevA.83.052704} {\bibfield  {journal} {\bibinfo  {journal} {Phys.
  Rev. A}\ }\textbf {\bibinfo {volume} {83}},\ \bibinfo {pages} {052704}
  (\bibinfo {year} {2011})}\BibitemShut {NoStop}%
\bibitem [{\citenamefont {Errea}\ \emph {et~al.}(2015)\citenamefont {Errea},
  \citenamefont {Illescas}, \citenamefont {Méndez}, \citenamefont {Rabadán},\
  and\ \citenamefont {Suárez}}]{ERREA2015}%
  \BibitemOpen
  \bibfield  {author} {\bibinfo {author} {\bibfnamefont {L.}~\bibnamefont
  {Errea}}, \bibinfo {author} {\bibfnamefont {C.}~\bibnamefont {Illescas}},
  \bibinfo {author} {\bibfnamefont {L.}~\bibnamefont {Méndez}}, \bibinfo
  {author} {\bibfnamefont {I.}~\bibnamefont {Rabadán}}, \ and\ \bibinfo
  {author} {\bibfnamefont {J.}~\bibnamefont {Suárez}},\ }\href {\doibase
  https://doi.org/10.1016/j.chemphys.2015.08.009} {\bibfield  {journal}
  {\bibinfo  {journal} {Chemical Physics}\ }\textbf {\bibinfo {volume} {462}},\
  \bibinfo {pages} {17 } (\bibinfo {year} {2015})}\BibitemShut {NoStop}%
\bibitem [{\citenamefont {{Otranto, Sebastian}}\ \emph
  {et~al.}(2019)\citenamefont {{Otranto, Sebastian}}, \citenamefont {{Bachi,
  Nicol\'as}},\ and\ \citenamefont {{Olson, Ronald E.}}}]{Otranto2019}%
  \BibitemOpen
  \bibfield  {author} {\bibinfo {author} {\bibnamefont {{Otranto, Sebastian}}},
  \bibinfo {author} {\bibnamefont {{Bachi, Nicol\'as}}}, \ and\ \bibinfo
  {author} {\bibnamefont {{Olson, Ronald E.}}},\ }\href {\doibase
  10.1140/epjd/e2019-90533-2} {\bibfield  {journal} {\bibinfo  {journal} {Eur.
  Phys. J. D}\ }\textbf {\bibinfo {volume} {73}},\ \bibinfo {pages} {41}
  (\bibinfo {year} {2019})}\BibitemShut {NoStop}%
\bibitem [{\citenamefont {Jorge}\ \emph {et~al.}(2020)\citenamefont {Jorge},
  \citenamefont {Horbatsch},\ and\ \citenamefont
  {Kirchner}}]{jorge2020multicharged}%
  \BibitemOpen
  \bibfield  {author} {\bibinfo {author} {\bibfnamefont {A.}~\bibnamefont
  {Jorge}}, \bibinfo {author} {\bibfnamefont {M.}~\bibnamefont {Horbatsch}}, \
  and\ \bibinfo {author} {\bibfnamefont {T.}~\bibnamefont {Kirchner}},\
  }\href@noop {} {\enquote {\bibinfo {title} {Multi-charged ion-water molecule
  collisions in a classical-trajectory time-dependent mean-field theory},}\ }
  (\bibinfo {year} {2020}),\ \Eprint {http://arxiv.org/abs/2001.06539}
  {arXiv:2001.06539 [physics.atom-ph]} \BibitemShut {NoStop}%
\bibitem [{\citenamefont {Kirchner}\ \emph {et~al.}(2001)\citenamefont
  {Kirchner}, \citenamefont {Horbatsch},\ and\ \citenamefont
  {L\"udde}}]{PhysRevA.64.012711}%
  \BibitemOpen
  \bibfield  {author} {\bibinfo {author} {\bibfnamefont {T.}~\bibnamefont
  {Kirchner}}, \bibinfo {author} {\bibfnamefont {M.}~\bibnamefont {Horbatsch}},
  \ and\ \bibinfo {author} {\bibfnamefont {H.~J.}\ \bibnamefont {L\"udde}},\
  }\href {\doibase 10.1103/PhysRevA.64.012711} {\bibfield  {journal} {\bibinfo
  {journal} {Phys. Rev. A}\ }\textbf {\bibinfo {volume} {64}},\ \bibinfo
  {pages} {012711} (\bibinfo {year} {2001})}\BibitemShut {NoStop}%
\bibitem [{\citenamefont {Faulkner}\ \emph {et~al.}(2019)\citenamefont
  {Faulkner}, \citenamefont {Abdurakhmanov}, \citenamefont {Alladustov},
  \citenamefont {Kadyrov},\ and\ \citenamefont {Bray}}]{Faulkner_2019}%
  \BibitemOpen
  \bibfield  {author} {\bibinfo {author} {\bibfnamefont {J.}~\bibnamefont
  {Faulkner}}, \bibinfo {author} {\bibfnamefont {I.~B.}\ \bibnamefont
  {Abdurakhmanov}}, \bibinfo {author} {\bibfnamefont {S.~U.}\ \bibnamefont
  {Alladustov}}, \bibinfo {author} {\bibfnamefont {A.~S.}\ \bibnamefont
  {Kadyrov}}, \ and\ \bibinfo {author} {\bibfnamefont {I.}~\bibnamefont
  {Bray}},\ }\href {\doibase 10.1088/1361-6587/ab2e7a} {\bibfield  {journal}
  {\bibinfo  {journal} {Plasma Physics and Controlled Fusion}\ }\textbf
  {\bibinfo {volume} {61}},\ \bibinfo {pages} {095005} (\bibinfo {year}
  {2019})}\BibitemShut {NoStop}%
\bibitem [{\citenamefont {Shah}\ and\ \citenamefont
  {Gilbody}(1981)}]{Shah_1981}%
  \BibitemOpen
  \bibfield  {author} {\bibinfo {author} {\bibfnamefont {M.~B.}\ \bibnamefont
  {Shah}}\ and\ \bibinfo {author} {\bibfnamefont {H.~B.}\ \bibnamefont
  {Gilbody}},\ }\href {https://doi.org/10.1088%2F0022-3700%2F14%2F14%2F009}
  {\bibfield  {journal} {\bibinfo  {journal} {Journal of Physics B: Atomic and
  Molecular Physics}\ }\textbf {\bibinfo {volume} {14}},\ \bibinfo {pages}
  {2361} (\bibinfo {year} {1981})}\BibitemShut {NoStop}%
\bibitem [{\citenamefont {Shah}\ \emph {et~al.}(1988)\citenamefont {Shah},
  \citenamefont {Elliott}, \citenamefont {McCallion},\ and\ \citenamefont
  {Gilbody}}]{Shah_1988}%
  \BibitemOpen
  \bibfield  {author} {\bibinfo {author} {\bibfnamefont {M.~B.}\ \bibnamefont
  {Shah}}, \bibinfo {author} {\bibfnamefont {D.~S.}\ \bibnamefont {Elliott}},
  \bibinfo {author} {\bibfnamefont {P.}~\bibnamefont {McCallion}}, \ and\
  \bibinfo {author} {\bibfnamefont {H.~B.}\ \bibnamefont {Gilbody}},\ }\href
  {https://doi.org/10.1088%2F0953-4075%2F21%2F13%2F012} {\bibfield  {journal}
  {\bibinfo  {journal} {Journal of Physics B: Atomic, Molecular and Optical
  Physics}\ }\textbf {\bibinfo {volume} {21}},\ \bibinfo {pages} {2455}
  (\bibinfo {year} {1988})}\BibitemShut {NoStop}%
\bibitem [{\citenamefont {Shah}\ and\ \citenamefont
  {Gilbody}(1982)}]{Shah_1982}%
  \BibitemOpen
  \bibfield  {author} {\bibinfo {author} {\bibfnamefont {M.~B.}\ \bibnamefont
  {Shah}}\ and\ \bibinfo {author} {\bibfnamefont {H.~B.}\ \bibnamefont
  {Gilbody}},\ }\href {https://doi.org/10.1088%2F0022-3700%2F15%2F3%2F023}
  {\bibfield  {journal} {\bibinfo  {journal} {Journal of Physics B: Atomic and
  Molecular Physics}\ }\textbf {\bibinfo {volume} {15}},\ \bibinfo {pages}
  {413} (\bibinfo {year} {1982})}\BibitemShut {NoStop}%
\bibitem [{\citenamefont {Rudd}\ \emph
  {et~al.}(1985{\natexlab{b}})\citenamefont {Rudd}, \citenamefont {Kim},
  \citenamefont {Madison},\ and\ \citenamefont {Gallagher}}]{Rudd85a}%
  \BibitemOpen
  \bibfield  {author} {\bibinfo {author} {\bibfnamefont {M.~E.}\ \bibnamefont
  {Rudd}}, \bibinfo {author} {\bibfnamefont {Y.~K.}\ \bibnamefont {Kim}},
  \bibinfo {author} {\bibfnamefont {D.~H.}\ \bibnamefont {Madison}}, \ and\
  \bibinfo {author} {\bibfnamefont {J.~W.}\ \bibnamefont {Gallagher}},\ }\href
  {\doibase 10.1103/RevModPhys.57.965} {\bibfield  {journal} {\bibinfo
  {journal} {Rev. Mod. Phys.}\ }\textbf {\bibinfo {volume} {57}},\ \bibinfo
  {pages} {965} (\bibinfo {year} {1985}{\natexlab{b}})}\BibitemShut {NoStop}%
\bibitem [{\citenamefont {Tabet}\ \emph
  {et~al.}(2010{\natexlab{a}})\citenamefont {Tabet}, \citenamefont {Eden},
  \citenamefont {Feil}, \citenamefont {Abdoul-Carime}, \citenamefont {Farizon},
  \citenamefont {Farizon}, \citenamefont {Ouaskit},\ and\ \citenamefont
  {M\"ark}}]{PhysRevA.81.012711}%
  \BibitemOpen
  \bibfield  {author} {\bibinfo {author} {\bibfnamefont {J.}~\bibnamefont
  {Tabet}}, \bibinfo {author} {\bibfnamefont {S.}~\bibnamefont {Eden}},
  \bibinfo {author} {\bibfnamefont {S.}~\bibnamefont {Feil}}, \bibinfo {author}
  {\bibfnamefont {H.}~\bibnamefont {Abdoul-Carime}}, \bibinfo {author}
  {\bibfnamefont {B.}~\bibnamefont {Farizon}}, \bibinfo {author} {\bibfnamefont
  {M.}~\bibnamefont {Farizon}}, \bibinfo {author} {\bibfnamefont
  {S.}~\bibnamefont {Ouaskit}}, \ and\ \bibinfo {author} {\bibfnamefont
  {T.~D.}\ \bibnamefont {M\"ark}},\ }\href {\doibase
  10.1103/PhysRevA.81.012711} {\bibfield  {journal} {\bibinfo  {journal} {Phys.
  Rev. A}\ }\textbf {\bibinfo {volume} {81}},\ \bibinfo {pages} {012711}
  (\bibinfo {year} {2010}{\natexlab{a}})}\BibitemShut {NoStop}%
\bibitem [{\citenamefont {Tabet}\ \emph
  {et~al.}(2010{\natexlab{b}})\citenamefont {Tabet}, \citenamefont {Eden},
  \citenamefont {Feil}, \citenamefont {Abdoul-Carime}, \citenamefont {Farizon},
  \citenamefont {Farizon}, \citenamefont {Ouaskit},\ and\ \citenamefont
  {M\"ark}}]{PhysRevA.82.022703}%
  \BibitemOpen
  \bibfield  {author} {\bibinfo {author} {\bibfnamefont {J.}~\bibnamefont
  {Tabet}}, \bibinfo {author} {\bibfnamefont {S.}~\bibnamefont {Eden}},
  \bibinfo {author} {\bibfnamefont {S.}~\bibnamefont {Feil}}, \bibinfo {author}
  {\bibfnamefont {H.}~\bibnamefont {Abdoul-Carime}}, \bibinfo {author}
  {\bibfnamefont {B.}~\bibnamefont {Farizon}}, \bibinfo {author} {\bibfnamefont
  {M.}~\bibnamefont {Farizon}}, \bibinfo {author} {\bibfnamefont
  {S.}~\bibnamefont {Ouaskit}}, \ and\ \bibinfo {author} {\bibfnamefont
  {T.~D.}\ \bibnamefont {M\"ark}},\ }\href {\doibase
  10.1103/PhysRevA.82.022703} {\bibfield  {journal} {\bibinfo  {journal} {Phys.
  Rev. A}\ }\textbf {\bibinfo {volume} {82}},\ \bibinfo {pages} {022703}
  (\bibinfo {year} {2010}{\natexlab{b}})}\BibitemShut {NoStop}%
\bibitem [{\citenamefont {Sarkadi}(2016)}]{Sarkadi_2016}%
  \BibitemOpen
  \bibfield  {author} {\bibinfo {author} {\bibfnamefont {L.}~\bibnamefont
  {Sarkadi}},\ }\href {\doibase 10.1088/0953-4075/49/18/185203} {\bibfield
  {journal} {\bibinfo  {journal} {Journal of Physics B: Atomic, Molecular and
  Optical Physics}\ }\textbf {\bibinfo {volume} {49}},\ \bibinfo {pages}
  {185203} (\bibinfo {year} {2016})}\BibitemShut {NoStop}%
\bibitem [{\citenamefont {Sarkadi}(2015)}]{PhysRevA.92.062704}%
  \BibitemOpen
  \bibfield  {author} {\bibinfo {author} {\bibfnamefont {L.}~\bibnamefont
  {Sarkadi}},\ }\href {\doibase 10.1103/PhysRevA.92.062704} {\bibfield
  {journal} {\bibinfo  {journal} {Phys. Rev. A}\ }\textbf {\bibinfo {volume}
  {92}},\ \bibinfo {pages} {062704} (\bibinfo {year} {2015})}\BibitemShut
  {NoStop}%
\bibitem [{\citenamefont {Rudd}\ \emph
  {et~al.}(1985{\natexlab{c}})\citenamefont {Rudd}, \citenamefont {Goffe},
  \citenamefont {DuBois},\ and\ \citenamefont {Toburen}}]{Rudd85c}%
  \BibitemOpen
  \bibfield  {author} {\bibinfo {author} {\bibfnamefont {M.~E.}\ \bibnamefont
  {Rudd}}, \bibinfo {author} {\bibfnamefont {T.~V.}\ \bibnamefont {Goffe}},
  \bibinfo {author} {\bibfnamefont {R.~D.}\ \bibnamefont {DuBois}}, \ and\
  \bibinfo {author} {\bibfnamefont {L.~H.}\ \bibnamefont {Toburen}},\ }\href
  {\doibase 10.1103/PhysRevA.31.492} {\bibfield  {journal} {\bibinfo  {journal}
  {Phys. Rev. A}\ }\textbf {\bibinfo {volume} {31}},\ \bibinfo {pages} {492}
  (\bibinfo {year} {1985}{\natexlab{c}})}\BibitemShut {NoStop}%
\bibitem [{\citenamefont {Ohsawa}\ \emph {et~al.}(2005)\citenamefont {Ohsawa},
  \citenamefont {Sato}, \citenamefont {Okada}, \citenamefont {Shevelko},\ and\
  \citenamefont {Soga}}]{Ohsawa2005}%
  \BibitemOpen
  \bibfield  {author} {\bibinfo {author} {\bibfnamefont {D.}~\bibnamefont
  {Ohsawa}}, \bibinfo {author} {\bibfnamefont {Y.}~\bibnamefont {Sato}},
  \bibinfo {author} {\bibfnamefont {Y.}~\bibnamefont {Okada}}, \bibinfo
  {author} {\bibfnamefont {V.~P.}\ \bibnamefont {Shevelko}}, \ and\ \bibinfo
  {author} {\bibfnamefont {F.}~\bibnamefont {Soga}},\ }\href {\doibase
  10.1103/PhysRevA.72.062710} {\bibfield  {journal} {\bibinfo  {journal} {Phys.
  Rev. A}\ }\textbf {\bibinfo {volume} {72}},\ \bibinfo {pages} {062710}
  (\bibinfo {year} {2005})}\BibitemShut {NoStop}%
\bibitem [{\citenamefont {Cappello}\ \emph {et~al.}(2009)\citenamefont
  {Cappello}, \citenamefont {Champion}, \citenamefont {Boudrioua},
  \citenamefont {Lekadir}, \citenamefont {Sato},\ and\ \citenamefont
  {Ohsawa}}]{DALCAPPELLO2009781}%
  \BibitemOpen
  \bibfield  {author} {\bibinfo {author} {\bibfnamefont {C.~D.}\ \bibnamefont
  {Cappello}}, \bibinfo {author} {\bibfnamefont {C.}~\bibnamefont {Champion}},
  \bibinfo {author} {\bibfnamefont {O.}~\bibnamefont {Boudrioua}}, \bibinfo
  {author} {\bibfnamefont {H.}~\bibnamefont {Lekadir}}, \bibinfo {author}
  {\bibfnamefont {Y.}~\bibnamefont {Sato}}, \ and\ \bibinfo {author}
  {\bibfnamefont {D.}~\bibnamefont {Ohsawa}},\ }\href {\doibase
  https://doi.org/10.1016/j.nimb.2008.12.010} {\bibfield  {journal} {\bibinfo
  {journal} {Nuclear Instruments and Methods in Physics Research Section B:
  Beam Interactions with Materials and Atoms}\ }\textbf {\bibinfo {volume}
  {267}},\ \bibinfo {pages} {781 } (\bibinfo {year} {2009})}\BibitemShut
  {NoStop}%
\end{thebibliography}%

\end{document}